\documentclass[acmlarge]{acmart}

\usepackage{amsmath,amsfonts,tabularx,pifont,makecell}
\usepackage{import}
\usepackage{multicol}
\usepackage{booktabs}
\usepackage{siunitx}
\usepackage{colortbl}
\usepackage{makecell}
\usepackage{multirow}
\usepackage{adjustbox}
\usepackage{enumitem}

\newcommand{\cmark}{\ding{51}}%
\newcommand{\xmark}{\ding{55}}%

\begin{document}

\title{Adversarial Challenges in Network Intrusion Detection Systems: Research Insights and Future Prospects}

\author{Sabrine Ennaji}
\email{ennaji@di.uniroma1.it}
\affiliation{%
  \institution{Sapienza University of Rome}
  \city{Rome}
  \country{Italy}
}

\author{Fabio De Gaspari}
\email{degaspari@di.uniroma1.it}
\affiliation{%
  \institution{Sapienza University of Rome}
  \city{Rome}
\country{Italy}}

\author{Dorjan Hitaj}
\email{hitaj.d@di.uniroma1.it}
\affiliation{%
  \institution{Sapienza University of Rome}
  \city{Rome}
  \country{Italy}}

\author{Alicia K bidi}
\email{kbidi@di.uniroma1.it}
\affiliation{%
 \institution{Sapienza University of Rome}
 \city{Rome}
 \country{Italy}}

\author{Luigi Vincenzo Mancini}
\email{mancini@di.uniroma1.it}
\affiliation{%
 \institution{Sapienza University of Rome}
 \city{Rome}
 \country{Italy}}

\renewcommand{\shortauthors}{Ennaji et al.}

\begin{abstract}
Machine learning has brought significant advances in cybersecurity, particularly in the development of Intrusion Detection Systems (IDS). These improvements are mainly attributed to the ability of machine learning algorithms to identify complex relationships between features and effectively generalize to unseen data. Deep neural networks, in particular, contributed to this progress by enabling the analysis of large amounts of training data, significantly enhancing detection performance. However, machine learning models remain vulnerable to adversarial attacks, where carefully crafted input data can mislead the model into making incorrect predictions. While adversarial threats in unstructured data, such as images and text, have been extensively studied, their impact on structured data like network traffic is less explored.

This survey aims to address this gap by providing a comprehensive review of machine learning-based Network Intrusion Detection Systems (NIDS) and thoroughly analyzing their susceptibility to adversarial attacks. We critically examine existing research in NIDS, highlighting key trends, strengths, and limitations, while identifying areas that require further exploration. Additionally, we discuss emerging challenges in the field and offer insights for the development of more robust and resilient NIDS. In summary, this paper enhances the understanding of adversarial attacks and defenses in NIDS and guide future research in improving the robustness of machine learning models in cybersecurity applications.

\end{abstract}
\maketitle



\section{Introduction}
The widespread deployment of internet-connected devices in our everyday lives has significantly impacted how we communicate, work, learn, and interact with the world. As society progresses in the information age, the Internet has evolved from a mere convenience to an indispensable necessity, playing a vital role in how our society functions, businesses operate, and people connect.
However, this high level of interconnectivity also creates opportunities for adversaries to exploit vulnerabilities in internet-facing services and devices for different purposes, including economic, political, or personal gain.
Some attacks exploit network protocol vulnerabilities~\cite{butler2009survey, aviram2016drown}, others focus on extracting and exploiting information through monitoring and sniffing network traffic~\cite{10.1007/978-3-319-11119-3_23, conti2018novel}, while others merely use the network as a medium to reach and misuse vulnerable systems and services~\cite{durumeric2014matter, antonakakis2017understanding}. 

Given the threat that cyberattacks present, improving the security of networks and computer systems has been a growing concern. Modern defenses typically rely on defense-in-depth approaches involving a variety of different strategies, such as firewalls~\cite{gouda2007structured}, cryptographic techniques~\cite{matta2021comparative}, host-based malware detection~\cite{gibert2021auditing}, autonomous defense systems~\cite{Theron2020, DeGaspari2019}, moving-target defense~\cite{jajodia2011moving, pagnotta2023dolos}, and Intrusion Detection Systems (IDS)~\cite{ozkan2021comprehensive}. Network Intrusion Detection Systems (NIDS), in particular, play a significant role in detecting various types of anomalous behavior due to their flexibility and ability to detect potential attacks before they reach a host by observing network traffic~\cite{can2015survey}. They are known to be efficient and have good potential to provide external and internal protection across services, devices, and architectures. In contrast to host-based IDS, NIDS do not need to run on each device that requires protection, but only require access to relevant network traffic. 
Despite these advantages, traditional NIDS based on expert knowledge require continuous updating to remain effective against evolving threats, as well as expert supervision to define detection rules. Over time, this can lead to performance degradation, reduced capability to detect novel attacks, and increased false positive rates. To address this gap, Machine Learning (ML) techniques and, more recently, Deep Learning (DL) have been increasingly integrated into NIDS to reduce the reliance on expert knowledge, improve generalization performance, and decrease the costs and complexity associated with updates~\cite{chou2021survey}. 
ML, and particularly Deep Neural Network (DNN), has revolutionized several domains due to the ability to extract valuable features from data and make accurate predictions. This is demonstrated by its success in areas like image classification~\cite{he2016deep, chollet2017xception}, natural language processing~\cite{nlp1, pagnotta2022passflow}, speech recognition~\cite{graves2013speech, speech3}, data generation~\cite{passGAN, karras2019style}, cybersecurity~\cite{hitaj2022maleficnet, 10158902, hitaj2024you}, forensics analysis~\cite{encod, de2020encod}, and even applications related to the COVID-19 pandemic~\cite{valenciacovidwired2021, lozano2021open}.

Critically, while extremely effective in controlled environments, ML techniques are susceptible to various attacks when exposed to adversarial settings~\cite{barreno2010security,michel2022survey}. Many attacks against ML have surfaced over the years, ranging from training-time attacks (e.g., backdooring,  poisoning~\cite{tian2022comprehensive, de2024have}) to inference-time evasion attacks (e.g., adversarial examples~\cite{goodfellow2015explaining, zhang2019adversarial}). Evasion attacks, in particular, present a fundamental threat to NIDS, as attackers possess considerable flexibility in modifying their traffic patterns to evade detection. Adversarial examples fall in this category of attacks, manipulating legitimate traffic data with subtle modifications to deceive the ML model and achieve an undesired output (incorrect classification). While adversarial examples were originally designed for unstructured data in the continuous domain~\cite{goodfellow2015explaining}, the significant risk they create spurred research investigating their applicability to other domains, such as network traffic, in an effort to respond to the question: 

\begin{center} \textit{Can adversarial attacks also threaten structured data?} \end{center}

\begin{figure}
  \centering
  \includegraphics[width=0.80\textwidth]{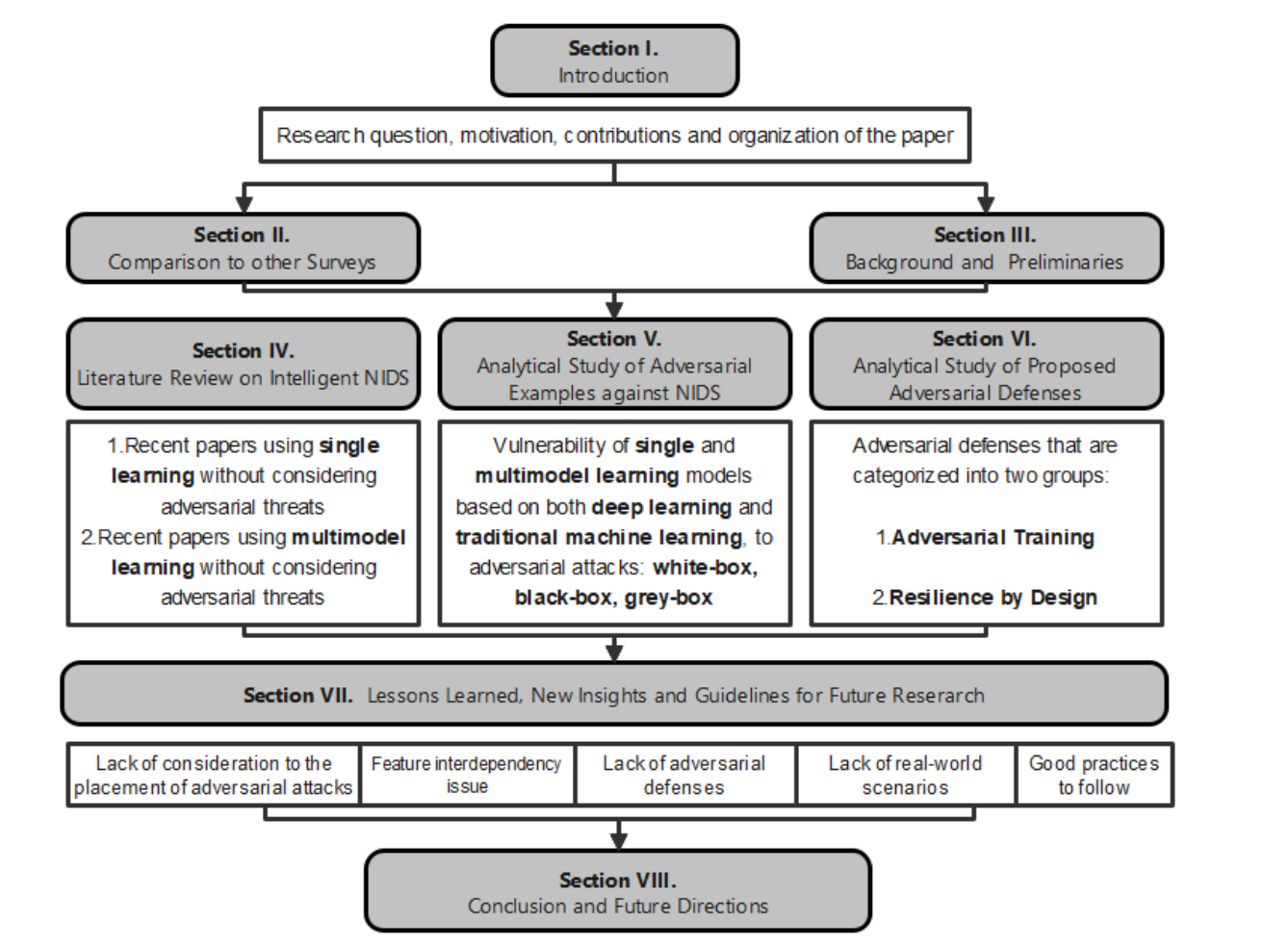}
  \caption{Structure of the Presented Survey}
\label{fig:structure}
\end{figure}

Recent studies~\cite{lin2022idsgan, rigaki2018bringing, wu2019evading, naked_sun, shu2020generative} demonstrate the applicability of evasion attacks and adversarial examples to structured data, introducing a significant vulnerability to cybersecurity applications, including ML-NIDS. However, existing research on NIDS design and evaluation often fails to adequately address this critical threat. This gap in knowledge motivates this work in three ways: 

\begin{itemize} [nosep]
   \item  \textbf{Raise awareness:} Currently, a limited number of studies about ML-NIDS include adversarial attacks in their threat models. Due to the difficulties in adhering to the constraints and interdependencies among features in structured data, these attacks are frequently overlooked. Furthermore, adaptive attacks designed to evade detection techniques remain only partially addressed in NIDS research. Given this potential risk to ML-NIDS, this paper aims to increase the awareness of these attacks in the NIDS community.
  \item \textbf{Provide a clear point of reference:} This developing field lacks comprehensive surveys on adversarial examples against NIDS. To the best of our knowledge, there is no existing survey that provides a full taxonomy of ML-NIDS, explores the interplay between adversarial examples and feature interdependencies within NIDS, and summarizes existing knowledge on evasion attacks and defenses for NIDS.
  We aim to bridge this gap and provide a new reference point for NIDS research, covering all aspects of robust network intrusion detection.
  \item \textbf{Establish best practices}: Existing studies on ML-NIDS use different datasets, evaluation metrics, and settings, which makes it difficult to compare findings and identify strengths and weaknesses within proposed solutions. This paper proposes a set of guidelines and best practices for consistent ML-NIDS evaluation, emphasizing the importance of testing against adaptive attacks.
\end{itemize}
\vspace{0.5em}
\textbf{CONTRIBUTIONS.} This survey provides a comprehensive, analytical review of ML-based NIDS and recent practical applications of adversarial attacks against NIDS. In summary, we make the following contributions:

\begin{enumerate} [topsep=0pt, partopsep=0pt, parsep=0pt, itemsep=0pt]
\item We review recent advancements in ML-based NIDS, providing a thorough taxonomy and analysis of existing approaches.
\item We compare and contrast different adversarial attacks against ML models, comprehensively discussing how they affect ML-based NIDS.
\item We discuss and critically review existing defenses against adversarial attacks on NIDS.
\item We provide an extensive analysis of lessons learned, novel insights on adversarial attacks on NIDS, and good practices and guidelines for future research.
\end{enumerate}
\vspace{0.5em}
\textbf{ORGANIZATION OF THE SURVEY.} 
Figure~\ref{fig:structure} outlines a schematic representation of this paper's structure, which is organized as follows.
Section~\ref{sec:comparison} reviews existing IDS surveys and highlights our contributions. Section~\ref{sec:background} provides general background on the evolution of NIDS, ML, DL, adversarial attacks, and their countermeasures. A comprehensive literature review of ML-NIDS is presented in Section~\ref{sec:nids_review}. Section~\ref{sec:nids_attacks} critically reviews literature on adversarial attacks against ML-NIDS, while Section~\ref{sec:defenses} expands on this to review possible defenses. Section~\ref{sec:lessons_learned} summarizes identified challenges and presents lessons learned, insights, and future research guidelines. Finally, Section~\ref{sec:conclusions} concludes our work.

\section{Comparison to Other Surveys}\label{sec:comparison}
Adversarial attacks against NIDS is an emerging research area lying at the intersection of cybersecurity and ML. While extensive surveys exist on NIDS, there is a lack of comprehensive analysis specifically focused on adversarial attacks against them. Existing studies fall into three categories: (1) surveys on ML and DL-based NIDS, (2) surveys on adversarial attacks and defenses in the continuous domain, and (3) surveys on adversarial attacks on malware detection. In the following sections, we critically analyze related works, identifying their limitations and highlighting the innovative aspects of our survey.

\subsection{Surveys on Traditional Machine Learning and Deep Learning Applications to NIDS}
Several surveys analyze the application of traditional ML and DL to NIDS. A foundational work in this area is by Buczak et al.~\cite{buczak2015survey}. They conducted a comprehensive literature study on ML for network intrusion detection around 2016. Their work presents a complete taxonomy of existing NIDS solutions and provides tutorial descriptions of each ML applied therein. In addition to an overview and description of commonly used datasets to train and test NIDS. Building upon this foundation, Chaabouni et al. \cite{chaabouni2019network} analyzed ML-NIDS techniques for securing Internet of Things (IoT) devices. Similarly to Buczak et al., they propose a thorough taxonomy of NIDS, refining it for the IoT context to address domain-specific problems and challenges. Ahmad et al.~\cite{ahmad2021network} furthered this progress by defining an updated taxonomy of ML-NIDS and critically reviewing existing techniques, highlighting challenges and proposing new research directions.
Many more general surveys on NIDS exist~\cite{drewek2021survey, thakkar2022survey, rupa2020review}, providing similar contributions and including the most recent ML-NIDS approaches proposed in the literature. However, while these references are valuable for new researchers in the field, they all share a crucial limitation: they fail to consider the problem of adversarial attacks to ML-NIDS and related countermeasures.

\subsection{Surveys on Adversarial Attacks and Defenses in the Continuous Domain}
Adversarial attacks against ML models in the continuous domain are a well-known challenge, both for traditional ML~\cite{barreno2010security} and DNN~\cite{zhou2022adversarial, xiong2023all}. Barreno et al.~\cite{barreno2010security} propose one of the first comprehensive analyses of the security of traditional ML techniques against adversarial attacks. In their work, the authors define a complete taxonomy of adversarial attacks against traditional ML algorithms and study the effectiveness of such attacks against practical applications such as spam filtering. In~\cite{zhou2022adversarial}, Zhou et al. survey existing adversarial attacks against DNN, both at training and inference time. The authors provide a comprehensive classification for such attacks and define a lifecycle for adversarial attacks and defenses. Xiong et al.~\cite{xiong2023all} propose a survey on the effects of different properties of training data on the adversarial robustness of DNN. The authors analyze several published works and propose a set of guidelines and techniques for improving adversarial robustness through data manipulation.

This line of surveys provides an invaluable resource for understanding all the existing constructions for adversarial attacks and defenses. However, all of them focus on the continuous domain (mainly images) and are not applicable to the context of NIDS, where complex interdependencies exist between features and constraints on feature values. Moreover, they typically do not consider challenges specific to cybersecurity and NIDS.

\subsection{Surveys on Adversarial Attacks in Malware Detection}

\noindent\textbf{General Malware Detection:}
Recently, researchers started studying the effects of adversarial examples in the area of malware detection. Gardiner et al.~\cite{gardiner2016security} studied the security of traditional ML techniques used for detecting malware's command and control channels. Their work systematically categorized existing command and control detection methods and analyzed the various attacks that could exploit these detectors.
In~\cite{park2020survey}, Park et al. provide a comprehensive survey of practical adversarial attacks against malware classifiers. Their paper explores various adversarial example-generation techniques, developing a framework to categorize them, and analyzes their strengths and weaknesses.
More recently, Rosenberg et al. ~\cite{rosenberg2021adversarial} propose in-depth survey on adversarial machine learning attacks and defenses specific to the cybersecurity field. Their study establishes a characterization of adversarial attacks based on three key aspects: lifecycle, adversarial goals, and attacker capabilities. Furthermore, they categorize different defense methods employed in cybersecurity applications.
These and other similar surveys (e.g., \cite{akhtar2021advances}, \cite{lan2022adversarial}) respond to the broader challenges of adversarial attacks on ML and robust malware detection. However, they lack a comprehensive analysis specific to NIDS and often focus on a particular area in adversarial attacks, neglecting the unique issues that arise within NIDS.

Finally, the analysis of potential defenses, when present, tends to be general and not specific to the context of malware detection.

\begin{table}
\small
\def\arraystretch{1.25}
\caption{Contributions of this survey in comparison to most the relevant previous works}
\centering
\begin{adjustbox}{max width=\textwidth}
\begin{tabular}[c]{c | c | c | c | c | c | c | c }
\hline
\rowcolor{lightgray}
\textbf{Reference} & \textbf{Year} & \textbf{\makecell{ML NIDS\\Taxonomy}} & \textbf{\makecell[c]{Feature\\Interdependency}} & \textbf{\makecell[c]{Placement of\\Adv. Ex.}} & \textbf{\makecell[c]{Real-World\\Challenges}} & \textbf{\makecell{NIDS Adv.\\Ex. Defenses}} & \textbf{\makecell{Lessons learned\\for NIDS}} \\
Ibitoye et al.~\cite{ibitoye2019threat} & 2019 & \xmark & \cmark & \xmark & \xmark& \cmark & \cmark \\
\hline
Olowononi et al.~\cite{olowononi2020resilient} & 2020 & \cmark & \xmark & \xmark & \xmark& \xmark & \xmark \\
\hline
Martins et al.~\cite{martins2020adversarial} & 2020 & \xmark & \cmark & \xmark & \cmark& \xmark & \xmark \\
\hline
Alotaibi et al.~\cite{alotaibi2023adversarial} & 2023 & \xmark & \xmark & \xmark & \cmark& \xmark & \xmark \\
\hline
Pinto et al.~\cite{pinto2023survey} & 2023 & \xmark &  \cmark &  \xmark & \cmark& \xmark & \xmark \\
\hline
He et al.~\cite{he2023adversarial} & 2023 & \cmark & \cmark & \xmark & \xmark& \cmark & \cmark \\
\hline
\hline
\textbf{Our survey} & \textbf{2024} & \cmark & \cmark & \cmark & \cmark& \cmark & \cmark \\
\end{tabular}
\end{adjustbox}
\label{tab:comparison}
\end{table}

\noindent\textbf{Intrusion Detection Systems:} In \cite{olowononi2020resilient}, Olowononi et al. review ML-based defenses for Networked Cyber-Physical Systems (CPS), examining recent intrusion detection techniques in CPS that leverage ML, with a focus on their robustness against adversarial attacks. Their study provides a full taxonomy of traditional ML and DNN intrusion detection methods for CPS and identifies promising future research directions. While the authors investigate the challenge of developing robust ML applications for intrusion detection, their focus lies in the distinct domain of CPS, separate from the area addressed by our survey.

Our paper provides a comprehensive taxonomy and in-depth analysis of ML-NIDS, identifying and exploring the unique challenges within this domain. Additionally, it explores methods specifically designed for adversarial attacks against NIDS, in contrast to \cite{olowononi2020resilient}, which focuses on general adversarial attacks. In \cite{martins2020adversarial}, Martins et al. survey applications of adversarial machine learning in intrusion and malware detection; they categorize general attack strategies and defenses before applying them to intrusion detection. Notably, their study lacks a general taxonomy of NIDS. Our survey distinguishes itself by presenting a detailed categorization of both ML-NIDS and the specific adversarial attacks that target them. We further present a thorough analysis of defenses against adversarial attacks for NIDS, which is an absent aspect in \cite{martins2020adversarial}.
Moreover, we discuss new insights, valuable lessons learned, and promising research directions in this field.
While in another thorough summary of several adversarial attacks and defenses for ML-NIDS explored by Ibitoye et al. \cite{ibitoye2019threat}, it does not provide a specific taxonomy of NIDS, but covers a wide variety of detectors such as spam detection, phishing detection, and general malware detection. Moreover, it does not consider the applicability of the discussed techniques. The presented defenses are also general and not specific to NIDS. Additionally, in \cite{alotaibi2023adversarial} an analysis of adversarial attacks with related defenses for ML-IDS is provided, but without discussing existing gaps in this area. A comprehensive taxonomy of ML-NIDS is not considered as well. 
Similarly, Pinto et al.~\cite{pinto2023survey} survey applications of ML to intrusion detection to protect critical infrastructure. A high-level categorization of ML-based defenses is introduced, but the challenge of adversarial examples is not thoroughly addressed. Finally, He et al.~\cite{he2023adversarial} propose a comprehensive analysis of existing works in ML-NIDS. They present a taxonomy of NIDS, a categorization of adversarial attacks and defenses against NIDS. In addition to an analysis of future research directions for this field. Compared to~\cite{he2023adversarial}, not only does our survey include additional recent papers in the literature, but we also critically review existing techniques, highlighting limitations and potential challenges to adoption. Moreover, we discuss new observations and correlations between the placement of adversarial examples within the data manifold~\cite{ma2018characterizing} and the resilience to adversarial attacks.

\subsection{Comparison Summary}
In summary, compared to related works, our survey is distinguished by comprehensive and methodical approach. We propose a complete taxonomy of ML-NIDS, adversarial attacks targeting them, and potential defenses. This analysis goes beyond simply listing approaches; we evaluate their strengths and weaknesses, providing summaries to guide practitioners. We further explore the real-world applicability of these attacks with their defenses in the NIDS context. Notably, our survey shed light on new insights on the existence and placement of adversarial examples and the interdependencies between malware features in NIDS. By analyzing these aspects and conducting a comprehensive review, we extract valuable lessons learned and propose a set of guidelines for developing robust ML-NIDS. Table~\ref{tab:comparison} summarizes the comparison of our survey to the most relevant related works and highlights significant gaps in existing surveys on adversarial examples against ML-NIDS. 
Surveys serve an indispensable purpose in consolidating knowledge and identifying research directions, which supports the advancement of a discipline. We believe our work offers detailed analysis and novel insights, making it a valuable resource for both new and experienced NIDS practitioners.
\section{Background}
\label{sec:background}
This section provides background information regarding intrusion detection systems, deep neural networks, common adversarial attacks on learning systems, and existing countermeasures.

\subsection{Intrusion Detection Systems}
Intrusion Detection Systems include network traffic monitoring (NIDS) and system behavior analysis (host-based IDS) to detect cyberattacks~\cite{aldweesh2020deep}. They work in conjunction with other security mechanisms, including firewalls, and use various detection techniques categorized into two main types: 

\begin{itemize}
    \item \textbf{Signature-based detection} uses information collected from previous attacks and known vulnerabilities to create a characteristic and recognizable signature. As attackers continuously update their attack patterns and methods, signatures become quickly obsolete and need to be regularly updated to remain effective~\cite{hubballi2014false}. Given its reliance on known attack patterns, signature-based detection typically offers high precision with low false positive rates but fails to detect new, unknown attacks. 

    \item \textbf{Anomaly-based detection} models the typical behavior of systems and networks and detects deviations from such behavior as potential attacks. The main advantage of this approach lies in its ability to identify both known and unknown attacks, as long as they cause a detectable behavioral shift. Furthermore, once the typical behavior of a system/network is computed, anomaly-based IDS do not need continuous updates. However, anomaly-based detection generally results in a higher rate of false positives compared to signature-based methods, making it more time-consuming to handle false alerts.
\end{itemize}

\subsection{Intelligent Intrusion Detection Systems}

\begin{figure}[bt]
  \centering
  \includegraphics[width=0.6\textwidth]{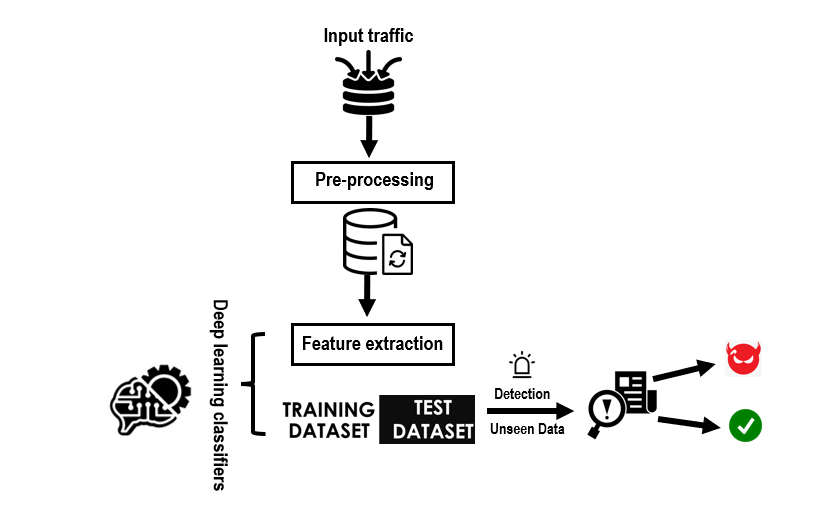}
  \caption{High-level architecture of a deep learning based-IDS}
  \label{fig:ids_architecture}
\end{figure} 

The growing complexity of cyberattacks led to significant advancements in IDS. In recent years, there has been a significant push towards integration of ML techniques in IDS, greatly enhancing detection capabilities~\cite{saranya2020performance}. A growing body of research highlights that ML consistently outperform traditional IDS methods in various settings~\cite{aldweesh2020deep}, with DL algorithms, in particular, showing promising results. This advantage arises from the ability of DL models to automatically learn and select descriptive features, eliminating the dependency on complex, expert-driven feature engineering (see Figure~\ref{fig:ids_architecture}) --- a major limitation in traditional ML~\cite{chiron2021fast}. Additionally, DL models excel at handling high-dimensional data and identifying intricate relationships between features, resulting in improved classification accuracy \cite{marin2018rawpower}. They also demonstrate strong adaptability to the dynamic nature of network traffic and evolving attack patterns. While DL training can be computationally expensive, advancements in software and hardware are mitigating this challenge~\cite{emani2022comprehensive}. Despite their numerous advantages, however, both ML- and DL-NIDS share a significant vulnerability: adversarial attacks~\cite{merzouk2022investigating, lin2022idsgan}. Adversarial attacks involve adversarial perturbation of training or query data designed to cause the model to fail in detecting malicious activity, thereby undermining its effectiveness.

\subsection{Deep Neural Networks}
Deep neural networks (DNNs) are algorithms developed to autonomously recognize, extract, and learn key patterns and relationships from input data, eliminating the need for extensive feature engineering by domain experts. DNNs excel at processing large volumes of high-dimensional data, and have consistently redefined state-of-the-art performance across various learning tasks. Practically, a DNN can be represented as a function (or a set of functions) that given an input provides an output as follows:

\begin{center}$f \in F : \mathbb{R}^n\rightarrow \mathbb{R}^m$\end{center}

Deep learning system comprises of two stages, the training and inference stages.

\begin{itemize}
 \item \textbf{The training stage:} Let $X$ be the class of inputs and $Y$ a label space. The learning procedure for a DNN model $f$ consists of the selection of a set of parameters $\theta$ that minimize a loss function $l$ on the training set $\mathcal{D=(X, Y)}$ which measures how well the DNN has learned over the data. The learning the parameters $\theta$ occurs through an iterative process where the loss function $l$, which measures the performance of the learned function $f$, provides the necessary information to optimize the parameters until an optimum is reached. More formally, neural network training seeks to optimize the following problem:

\begin{center}$ \underset{\theta}{argmin} \sum_{(x,y)\sim \mathcal{D}} l(\theta, x, y)$ \end{center}

where $\theta$ are the parameters of the DNN to optimize, $x$ and $y$ are the training datapoints and their corresponding labels sampled from the training data $\mathcal{D}$, and $l$ is the loss function used for training.

 \item \textbf{The inference stage:} After finding an optimal set of parameters $\theta$, the model can be used to classify new data. An unseen input $x_j$  and its corresponding label $y_j$ are calculated as follows:

\begin{center}$ y_j = f(\theta, x_j) $\end{center}

Additionally, during the classification task, the output of the last layer $Y$ (called \textit{logits}) are transformed into a probability distribution by an activation function (e.g., Softmax):

\begin{center}$ 0 \leq Y_i \leq 1 \wedge Y_1+...+Y_m=1; $\end{center}
where $Y_i$ can be understood as the probability that input $x$ has class $i$. The label that has the highest probability $argmax (Y_i)$ presents the class of the input $x$.
\end{itemize}

\subsection{Adversarial Threats}\label{sec:adversarial_examples}
Traditional ML and DL techniques have been show to be vulnerable to adversarial attacks: manipulations of input data that lead models to produce wrong predictions. One particularly concerning form of such attacks are \textit{adversarial examples}~\cite{szegedy2014intriguing, goodfellow2015explaining}. Initially studied in image classification task, adversarial examples involve subtle, often imperceptible perturbations to the input data that can significantly alter DNN's output. 
Generally, the training process of a DNN involves minimizing a loss function $l$, which quantifies the error between the target and the predicted classes. This optimization process aims to reduce the error rate, improve the model's accuracy by adjusting its parameters based on the computed loss. The overall error rate used to evaluate the model's performance is then calculated by comparing the sum of the loss between the target and the corresponding labels.

 \begin{center}$ \sum_{(x,y)\sim \mathcal{D}} l(f(\theta,x),y) $\end{center}

Adversarial examples are crafted by adding a small \textit{adversarial perturbation} to a clean input $x$. The details on how the perturbation is generated depend on the specific algorithm $\mathcal{A}$ used, but for DNNs, they typically rely on gradient computation to increase the loss (i.e., error) of the model on a specific input $x$: 

\begin{center} $x_{adv} = (x+\mathcal{A}(\theta,x,y))$ \end{center}

Over the years, numerous adversarial attacks have been introduced, and they can be categorized based on four main criteria:

\begin{enumerate}

\item \textbf{Attack Timing:} attacks are categorized according to when they happen. Poisoning attacks exploit the model's vulnerability during the training phase by injecting manipulated data in the training set, aiming to corrupt the learning process~\cite{tian2022comprehensive}. Evasion attacks, such as adversarial examples, occur during the test phase, where adversarial perturbations are introduced into the query sample to deceive the model into making incorrect predictions~\cite{yang2021rumor, eykholt2018robust, martins2020adversarial}. 

\item \textbf{Adversarial Goal:} adversarial attacks can be either targeted or untargeted. Targeted attacks aim to misclassify the input as a specific incorrect class. Untargeted attacks aim to force the model to misclassify the input, without specifying which class it should be classified as.

\item \textbf{Adversary's Knowledge:} adversary capabilities are classified based on the level of access to information about the target model.
\begin{itemize}
\item \textit{White box-attacks:} the attacker has full knowledge about the model (architecture, training data, model parameters) and can exploit this information to craft highly effective adversarial examples. Many white-box adversarial attacks generation methods exist, such as Limited-memory Broyden–Fletcher–Goldfarb Shanno~\cite{szegedy2014intriguing}, Fast Gradient Sign Method~\cite{goodfellow2015explaining}, Basic Iterative Method, Projected Gradient Descent~\cite{kurakin2018adversarial}, Deepfool~\cite{moosavi2016deepfool}, Carlini \& Wagner~\cite{carlini2017towards}, and Jacobian-based Saliency Map Attack~\cite{linardatos2020explainable}.
\item \textit{Black-box attacks:}  the attacker has no knowledge of the model's architecture, parameters, or training data, and can only interact with it in an oracle-like fashion by querying specific inputs and observing the corresponding outputs. Consequently, black-box attack typically rely on iterative methods, where the adversarial sample is updated over time to achieve the desired adversarial goal. These attacks often employ techniques like Generative Adversarial Networks (GAN)~\cite{goodfellow2016nips, martins2020adversarial} or Zeroth-Order Optimization attacks~\cite{chen2017zoo}, but can also utilize on surrogate models by approximating the target model’s behavior and then use these learned models to generate adversarial inputs for the original system~\cite{zhou2020dast}. 
\item \textit{Grey-box attacks:} 
the attacker has partial knowledge about the DNN model. Unlike black-box access, where only the inputs and outputs of the model are observable, grey-box access allows some limited understanding or control over the model's internal structure or behavior. This can include access to certain layers or parameters but not full transparency like white box access, where the entire model's structure and training process are fully known and accessible.
\end{itemize}

\item \textbf{Attack Frequency:} 
adversarial attacks can be one-shot attacks or iterative attacks. One-shot attacks generate adversarial perturbations are through a single-step algorithm. Iterative adversarial attacks rely on multiple iterations, repeatedly querying the classifier to refine the adversarial perturbation.
\end{enumerate}

\subsection{Adversarial Countermeasures}\label{sec:adversarial_examples_countermeasures}
Given the relevance of adversarial examples and the threat they pose to the integrity of ML and DL applications, several defense mechanisms have been explored over the years. Several strategies have been explored, such as 

adversarial training, model regularization, input preprocessing, and robust optimization, aiming to enhance the resilience and reliability of neural networks against such threats.
Despite much research effort, however, most proposals have proven ineffective and, over time, have succumbed to novel adaptive attacks. Figure~\ref{fig:adv_ex_timeline} presents the timeline of these defenses and corresponding adaptive attacks that reveal their limitations~\cite{carlini2017adversarial, carlini2017magnet, athalye2018obfuscated, carlini2019evaluating, tramer2020adaptive}.
Adversarial example countermeasures can be divided into two main categories: 

\begin{figure}[t]
  \centering
  \begin{adjustbox}{max width=\textwidth}
    \includegraphics{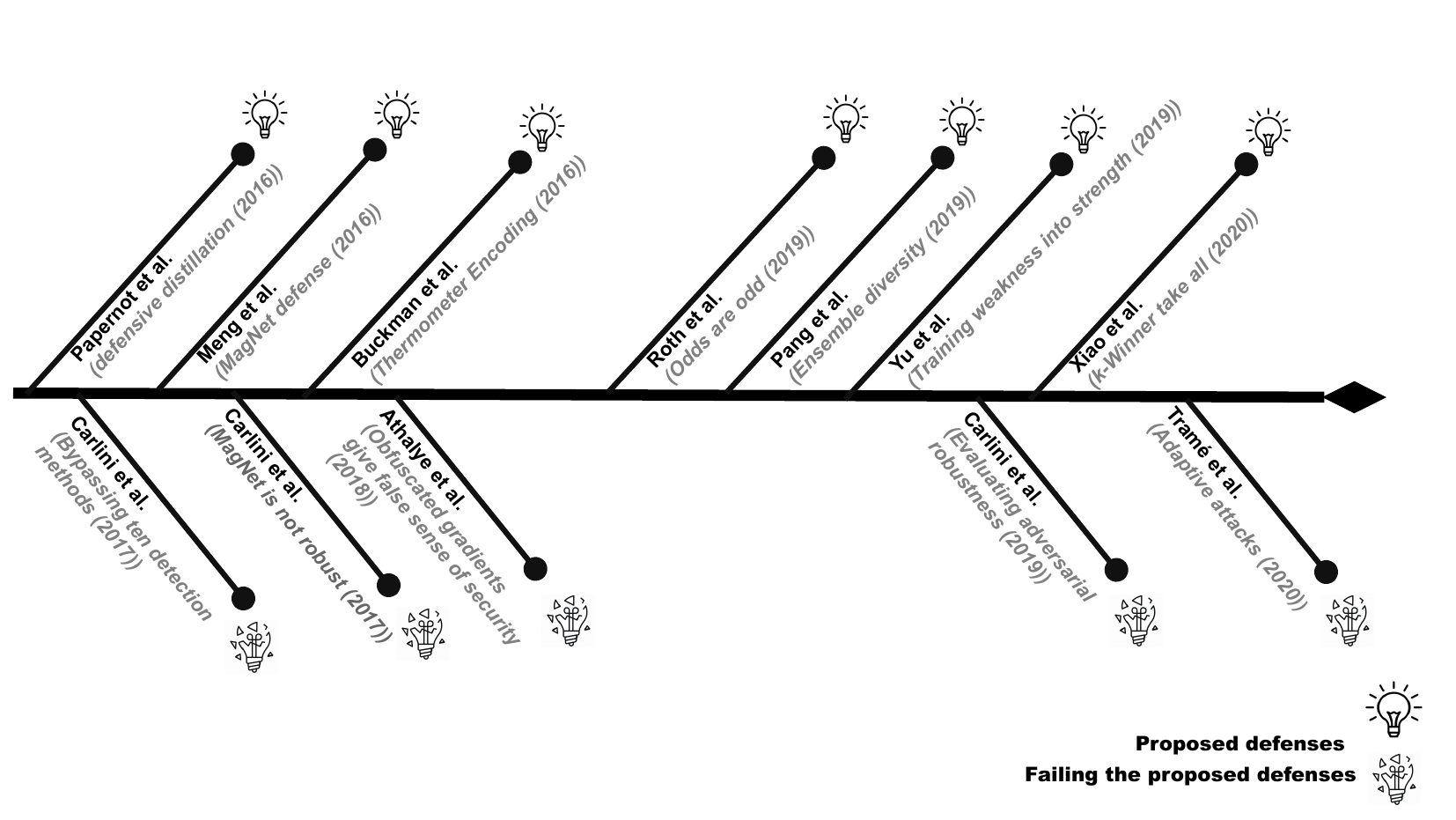}
  \end{adjustbox}
  \caption{Timeline of adversarial attacks and defenses.}
  \label{fig:adv_ex_timeline}
\end{figure}

\begin{enumerate}
\item \textbf{Adversarial Training:} the goal of adversarial training approaches is to enable the model to develop a robust representation of the training data, thereby reducing the impact of adversarial examples. This is accomplished by training the model using both clean and adversarial datapoints, allowing for an adaptation of the learned decision boundary to accurately classify both perturbed and unperturbed instances~\cite{goodfellow2015explaining}. Formalizing, adversarial training aims to find a set of model parameters $\theta$ that minimize the following expression:

\begin{center}$\underset{\theta}{argmin}\; E_{(x,y)\sim D} [\underset{\delta \in \Delta}{max} \; l(\theta, x+\delta, y)] \; \; \; \;  (4) $ \end{center}

where $\delta$ is an adversarial perturbation, $\Delta$ represents the set of all allowable adversarial perturbations, $D$ denotes the data distribution over inputs $x \in \mathbb{R}^d$ and labels $y$, $l(\theta, x, y)$ is the loss function for the model parameters $\theta$.

\item \textbf{Architecture Reinforcing:}
this category of countermeasures includes several methods that share the common purpose of reinforcing the model's architecture, each employing different strategies as detailed below.

\begin{itemize}
\item \textit{\textbf{Detection and Rejection:}} this method involves using trained detectors to evaluate model inputs and detect potential adversarial perturbations. When a perturbation is detected, the system executes a series of specialized operations to reduce/eliminate the identified threat \cite{metzen2017detecting}.

\item \textit{\textbf{Input Transformation:}} this approach applies a set of transformations to the input data to eliminate potential adversarial perturbations before the sample is used to query the model. The input transformations typically aim at disrupting the features altered by adversarial perturbations, mitigating the effect of the attacks~\cite{ali2022evaluating, bhagoji2018enhancing}.

\item \textit{\textbf{Gradient Masking:}} this technique relies on hiding the internal state of the model --- its gradient on specific inputs. Since most adversarial example algorithms require gradient computation, the attacks are ineffective if gradient information cannot be obtained accurately~\cite{chen2022adversarial}.

\item \textit{\textbf{Manifold Projection:}} this approach leverages the natural data distribution (manifold) to detect and remove adversarial perturbations~\cite{ma2018characterizing, stutz2019disentangling}. Adversarial examples typically reside in different region of the domain compared to clean samples, and 
manifold projection can employed to discard adversarial examples that deviate significantly from the natural data structure.
 
\item \textit{\textbf{Ensemble Method:}} 
this approach relies on the combination of multiple models and model diversity to make it more challenging for attackers to generate effective adversarial examples~\cite{tramer2018ensemble, pang2019improving}.

\end{itemize}
\end{enumerate}
\section{Literature Review of Intelligent Network Intrusion Detection Systems}
\label{sec:nids_review}

Machine learning, especially deep learning, shows promise in NIDS. While traditional methods like Support Vector Machines (SVM), Decision Trees (DTs), and Naive Bayes (NB) are simpler and easier to interpret, they perform poorly against complex attacks. DL-NIDS are more effective but require more resources and are less transparent in their decision-making. This section presents a detailed review of both traditional and DL-NIDS, categorized by learning approach (single vs. multimodal).
This is summarized comprehensively in Table~\ref{tab:lit_review}.

\subsection{Single Learning}
\label{sec:single_learning}
Single Learning refers to a machine learning approach where a single, monolithic model is used to perform anomaly detection or classification on unseen data. 

\subsubsection{Traditional Machine Learning Approaches:} Research persists on traditional ML methods for NIDS due to their interpretability compared to DL, aiming to bridge the performance gap with DNN. Chkirbene et al.~\cite{chkirbene2020tidcs} developed two NIDS models based on a hybrid feature selection approach: Trust-based Intrusion Detection and Classification System (TIDCS), and an accelerated variant (TIDCS-A). Their method combines randomization, wrapper, and filter strategies to identify the most descriptive features to use for classification. First, TIDCS randomly selects multiple subsets of features and trains a Random Forest (RF) classifier on each subset. Then, the trained models are used to compute the score of individual features in each subset based on the accuracy achieved by the classifier. Finally, the best-scoring features across all subsets are selected for training the final model. Despite achieving good classification results, the reliance on pre-trained model accuracy for feature selection can introduce bias in the classification. Moreover, focusing only on increasing accuracy during feature selection can lead to decreased robustness, making the model more susceptible to evasion attacks.

Alahmadi et al.~\cite{alahmadi2020botection} propose a bot identification method that combines Markov Chains and RF classifiers. By modeling network behavior as state transitions in the Markov Chain, they capture high-level protocol states as features for botnet family detection. This focus on high-level behavior enhances robustness, as modifying the malware logic would be necessary for evasion. However, similar techniques based on system call transition analysis have been demonstrated to remain vulnerable to evasion attacks~\cite{nunes2022bane}. Moreover, potential overfitting on specific network patterns and the lack of cross-dataset validation require further investigation.
For detecting Denial of Service (DoS) attacks in Software-Defined Networks (SDN), Ramamurthy et al.~\cite{mehr2019svm} suggest an in-controller NIDS that leverages control packets generated by SDN when encountering previously unseen network flows. The proposed NIDS analyzes packets' data distribution, such as IP addresses, ports, and timing, using an entropy metric for anomaly detection with an SVM model. Although this transparent method simplifies SDN integration, its reliance on easily manipulated features makes it susceptible to evasion attacks.

\textbf{TAKEAWAY.} Traditional ML in NIDS often requires complex and sophisticated feature selection and dimensionality reduction for optimal performance. Although this increases the complexity of the NIDS architecture, it enables a more focused classification on specific aspects of the network traffic. However, complex designs may result in biases and poor generalization to unseen data. Moreover, although selected features are often descriptive, they generally lack resilience to adversarial attacks due to the challenges in creating robust selection and reduction strategies.

\subsubsection{Deep Learning Approaches:} To address limitations of traditional ML, recent NIDS research prioritizes DL. This focus employs DNN's ability to learn complex features from data, aiming for better classification. This focus can be divided into two categories: fully unsupervised feature representation and guided feature representation.

\vspace{0.5em}\noindent\textit{\textbf{Fully Unsupervised Feature Representation.}}
This family of works uses end-to-end DNN to autonomously learn feature representations from unlabeled data and perform classification. In~\cite{zhou2018deep}, the authors developed DFEL, a NIDS based on a DL approach for feature embedding learning. They pre-trained a DNN on historical NIDS data (normal and abnormal traffic), using it as a feature extractor for IoT traffic. These extracted features were then used to train another DNN for anomaly detection, leveraging the generalist DNN's data representation to classify IoT network traffic.

Similarly, Manimurugan et al.~\cite{manimurugan2020effective} use a Deep Belief Network (DBN) to learn descriptive features from raw network traffic. They combine unsupervised pre-training with supervised fine-tuning to detect IoT network attacks. The DBN learns effective feature representations during the unsupervised phase, which are then used for classification in the fine-tuned model. 
In line with~\cite{zhou2018deep, manimurugan2020effective}, Alom et al.~\cite{alom2017network} developed a DNN-NIDS based on neuromorphic computing systems by combining unsupervised and supervised training to learn descriptive feature representations and perform classification.  An AE is used for unsupervised feature learning from unlabeled data, which is then fed to a fully-connected network trained on labeled data. The trained DNN's weights are quantized and mapped to an IBM TrueNorth neuromorphic SoC for evaluation.

Fully unsupervised feature representation approaches enable the utilization of available unlabeled data --- such as general network traffic --- to learn useful features for similar domains with limited labeled data (e.g., IoT), ultimately improving the performance of the final classifier. However, relying on fully unsupervised feature learning risks overfitting on characteristics that are specific to the source domain but absent in the target domain, introducing biases in the final model. Furthermore, exclusively relying on unsupervised feature learning increases vulnerability to adversarial attacks, as unconstrained (or semi-constrained) features can be more easily exploited for evasion ~\cite{pawlicki2020defending}.

\vspace{0.5em}\noindent\textit{\textbf{Guided Feature Representation.}}
This family of works uses a combination of DNN and other techniques to learn descriptive traffic embeddings for intrusion detection. Laghrissi et al. \cite{laghrissi2021intrusion} combine Long Short-Term Memory (LSTM) with Principal Component Analysis (PCA) and Mutual Information (MI) for dimensionality reduction and feature selection. The approach first uses PCA to transform correlated features into a smaller set of uncorrelated features. Then, MI is employed to compute the statistical dependence between individual features and their labels, selecting those highly relevant to the target class. The model is then trained on these features for both binary and multiclass classification.
In~\cite{kan2021novel}, Kan et al. developed a CNN-NIDS using adaptive particle swarm optimization (APSO) to tune the model's hyperparameters. They use a custom network dataset from various IoT devices and manually select aggregate statistical features of the traffic (e.g., correlation coefficients on packet sizes, mean, variance, etc.). The feature vector is then fed to a 1-dimensional CNN for feature learning and classification. APSO is used to optimize the CNN's hyperparameters based on a held-out validation set to evaluate the fitness of the offsprings.
Thakkar et al.~\cite{thakkar2023fusion} use a fusion-based method on statistical features derived from raw traffic data. Features for training are selected based on a ranking system considering standard deviation, mean, and median. They are added iteratively in ranking order until the model's accuracy no longer improves. Piskozub et al. propose MalPhase in~\cite{piskozub2021malphase}, a multi-phase technique for anomaly detection, malware type classification, and malware family classification using augmented NetFlow data. It combines expert knowledge for feature selection with AE-based feature representation learning to extract descriptive features. The AE is trained to reconstruct original malicious data by combining benign and malicious traffic, forcing it to learn core features of malware traffic. MalPhase uses a multi-tiered architecture: lower-tier models analyze short time windows for brief spikes in activity, while higher-tier models identify sustained malware communication. In~\cite{zhong2020helad}, Zhong et al. combine had-crafted feature engineering, unsupervised feature learning, and supervised DNN-based classification to identify malicious network traffic. Damped Incremental Statistics~\cite{mirsky2018kitsune} is used to extract features from flow entries of SDN switches, and a fully connected AE is used to further refine the feature representation. The trained autoencoder is used to obtain an ``anomaly score'' and identify potential anomalous traffic. The anomalous traffic is then used to train an LSTM network and classify unseen network traffic as either normal or malicious.

The main advantage of combining DL with statistical and manual feature selection techniques lies in the integration of DNN's ability to identify non-linear correlations in data with domain-expert knowledge. This combination aids in identifying the most descriptive features, while leveraging prior expert knowledge to mitigate biases in the training data. Furthermore, combining expert knowledge with DNN for feature representation can enhance robustness against evasion attacks, provided that the features are well-designed to limit manipulation. However, with a partial exception of~\cite{piskozub2021malphase}, none of the previously mentioned works consider robustness in their feature representation and selection designs. Additionally, their threat models do not account for sophisticated evasion strategies, such as adversarial examples or adaptive attacks.

\textbf{TAKAWAY.} DL-NIDS show several advantages over traditional ML-NIDS. DNNs can reduce the need for human intervention by learning unsupervised feature representations. They can also identify complex, non-linear relations in the training data, which typically results in improved performance compared to traditional ML techniques. Despite adversarial noise, unsupervised DNN techniques can be leveraged to learn robust embeddings from unlabeled data~\cite{piskozub2021malphase}, which improves robustness. However, DL-NIDS are susceptible to adversarial attacks that adversaries can exploit to compromise their performance, limiting their reliability. While some approaches are designed to be robust against simpler attacks, such as noise injection, resilience against sophisticated adversarial attacks is generally lacking. Moreover, DNN-NIDS face challenges due to their black-box nature, as they lack formal methods to analyze their decision-making process. While this drawback can be mitigated by incorporating human expert supervision in the feature selection process, it cannot be entirely eliminated.

\subsection{Multimodel Learning}
\label{sec:multimodel_learning}
In contrast to single learning, multimodel or ensemble learning combines the prediction of multiple models to improve performance and resilience, reducing the error rate if model errors are independent and uncorrelated.

\subsubsection{Combination of Multiple Traditional Machine Learning Algorithms: }
Several works propose combinations of different traditional ML techniques to address the limitations of any individual approach. Tama et al.~\cite{tama2019tse} propose a feature selection and ensemble classification method for intrusion detection. They employ evolutionary algorithms like particle swarm optimization to identify the most descriptive features. These algorithms evaluate feature subsets based on the classification accuracy of a decision tree. Next, a two-stage ensemble classifier is utilized: a rotation forest acts as a filter, selecting feature subsets for training a bagged ensemble of weak learners. Finally, majority voting combines individual predictions for the final classification.
Similarly, Zhou et al.~\cite{zhou2020building} suggest a two-stage approach with feature selection followed by ensemble classification. They employ Correlation-Based Feature Selection for initial selection and an evolutionary algorithm to eliminate redundant features. However, unlike Tama et al.~\cite{tama2019tse}, they use a decision tree and random forest ensemble for classification. 
While the adoption of an ensemble of weaker classifiers is typically more resilient to changes in data distribution and tends to generalize better than single-model approaches~\cite{dong2020survey}, the overreliance of these approaches on performance-based metrics for feature selection is likely to introduce biases tied to the training data, ultimately compromising generalization capabilities. 
 
In \cite{gu2019novel}, an intrusion detection method combining an ensemble of SVM models with feature augmentation and train set splitting is proposed. The authors use fuzzy-c means clustering to divide training data into overlapping subsets, which are then processed and augmented with multivariate marginal density ratio features. An ensemble of SVM classifiers is trained on the augmented subsets, and their outputs are combined by a separate SVM model for the final classification. Despite achieving good performance on the considered dataset, relying on ensembles of homogeneous models can lead to decreased robustness to adversarial attacks compared to heterogeneous ensembles~\cite{pang2019improving}. 

\textbf{TAKAWAY.} 
Ensembles of traditional ML algorithms help improve performance and can sometimes improve generalization to data with different distributions, especially when heterogeneous models are used. However, the use of ensembles leads to increased computational cost and complexity, limiting its applicability to certain settings. Excessive reliance on performance metrics in feature selection can introduce biases, which may hinder the model's ability to generalize effectively to new data. Lastly, while less complex, homogeneous ensembles are generally more vulnerable to adversarial attacks~\cite{pang2019improving}.

\subsubsection{Combination of Deep Learning and Traditional Machine Learning: }
Due to the limitations of single-model DL techniques and ML-based approaches, several studies propose hybrid models that combine DL and traditional ML to compensate for each other's weaknesses.
Marir et al. \cite{marir2018distributed} proposed a distributed framework for large-scale network intrusion detection that combines a Deep Belief Network (DBN) with a multilayer ensemble SVM. The DBN is utilized as a feature extractor, and the ensemble of SVM is used to classify samples based on majority voting. Although the framework performs well in a variety of network settings, the use of multiple, multi-layer networks is resource-intensive and complex to maintain. Moreover, it is challenging to tune the hyperparameters of the networks without introducing biases tied to the training data.
In~\cite{imran2022intelligent}, Imran et al. combine a non-symmetrical deep auto-encoder (NDAE) with an SVM for network traffic classification. They use an encoder-only architecture to encode input data into latent representations, omitting reconstruction, which decreases computational time and costs. Similar to other works leveraging AEs~\cite{piskozub2021malphase, marir2018distributed}, the proposed approach demonstrates an improved ability to learn complex feature correlations and identifying significant features. However, the proposed architecture does not account for robustness to adversarial examples and adaptive attacks, and its heavy reliance on unguided, fully unsupervised learning risks overfitting on the specific characteristics of the training data. 
Ravi et al.~\cite{ravi2022recurrent} propose a recurrent neural network (RNN)-based feature fusion method consisting of two phases: (i) feature learning using an RNN network and dimensionality reduction, and (ii) classification with an heterogeneous ensemble of RF, SVM, and logistic regression model. First, the RNN feature extraction captures temporal correlations in the input data and Kernel Principal Component Analysis reduces the dimensionality of the extracted features. Finally, the classification is carried out by the ensemble through a voting mechanism. While effective, this complex classification pipeline is computationally intensive and challenging to utilize in real-time scenarios. Additionally, features based on temporal correlations are particularly vulnerable to adversarial attacks, making them an easy target for manipulation~\cite{afianian2019malware}.
Recently, Arreche et al.~\cite{arreche2024two} introduced a two-level ensemble learning framework. The first level of the framework uses base an ensemble of models to generate both a class prediction and a prediction probability. The second level of the framework combines the predictions and the probabilities from the first level models to further refine the classification through a different ensemble of models. The framework employs a variety of model architectures (DNN, DT, KNN, SVM) and boosting techniques for the two ensembles, creating a highly heterogeneous architecture. While the approach shows good performance across several dataset, the framework is computationally intensive, limiting its practicality in real-time scenarios. Furthermore, the complex interplay among the numeours classifiers across the two levels of the framework makes it difficult to explain the decision-making process, and may introduce overfitting and subtle biases tied to the training data.

\textbf{TAKAWAY.} 
Heterogeneous ensembles combining DL and traditional ML models are generally effective for intrusion detection. DNNs enable the effective use of large (unlabeled) datasets and serve as powerful feature extractors, while traditional ML algorithms provide accurate classification with lower computational requirements. However, relying on fully unsupervised feature learning may lead to increase susceptibility to adversarial attacks, as well as potentially introducing biases tied to the training data. Moreover, the increased architectural complexity of large ensembles introduces restrictions in the applicability of the techniques in resource- or time-constrained scenarios. Finally, the reliance on DL for feature learning and extraction compromises the explainability provided by traditional ML techniques, making it more difficult to interpret the model's decision-making process.

\subsubsection{Combination of Multiple Deep Learning Algorithms:}
To leverage the ability of DNNs to autonomously learn descriptive features and identify non-linear correlations in data, several studies have proposed combining multiple DL models to enhance detection performance.
Mirsky et al~\cite{mirsky2018kitsune} proposed Kitsune, an AE-based ensemble NIDS. Kitsune is a fully unsupervised approach that leverages Damped Incremental Statistics and AE reconstruction error rate to detect potentially anomalous traffic. The approach computes temporal statistics for each communication channel in the network, which are then mapped and input to a set of autoencoders. The reconstruction error of each AE is further processed by another fully connected AE to produce a final anomaly score for the traffic. While the completely unsupervised nature of Kitsune makes it easier to train and deploy, the reliance on temporal features makes it susceptible to evasion attacks. Additionally, the fully unsupervised nature of the approach further exacerbates this vulnerability.
Nandanwar et al.~\cite{nandanwar2024deep} have recently proposed AttackNet, a model based on the combination of CNN and Gated Recurrent Units (GRU) for classifying network time-series data.
The proposed architecture utilizes a limited number of gates to mitigate overfitting. Furthermore, the combination of GRU and CNN enables the analysis of both temporal and spatial features of data, improving the detection capabilities of the ensemble. Hasan et Securing industrial
internet of things against botnet attacks using hybrid deep learning approachal.~\cite{hasan2022securing} tackle intrusion detection within IIoT environments through a hybrid DL strategy combining LSTM and DNN. An LSTM netowkr processes the time series sensor data, while a DNN separately processes non-time series traffic features. A single dense layer is used to combine the outputs of the LSTM and DNN and perform classification. Similarly, a recent study by Nazir et al.~\cite{nazir2024deep} explores the application of LSTM and CNN to anomaly detection in IoT networks. Initially, a CNN is used to learn and extract structural features of the data, followed by a pooling layer to reduce the dimensionality of these features. In parallel, the input data is fed to an LSTM network to identify temporal correlations. The outputs of the LSTM and CNN are combined through a dense layer to perform classification. All these approaches utilize a similar detection architecture, heavily relying on temporal correlations in the network traffic to identify malicious activity. While temporal relations are powerful features for classifying in-distribution data, their unconstrained nature makes them vulnerable to manipulation by attackers, allowing for easier evasion of detection.
Yaser et al.~\cite{yaser2022improved} introduce a two-step architecture to detect DDoS attacks based on AE, CNN, and LSTM networks in SDN environments. A fully connected AE is used to learn descriptive features from the flow table entries and associated statistics of SDN switches. A CNN and an LSTM network are then trained on the labeled features extracted by the autoencoder to identify DDoS traffic. While unsupervised feature learning enables the extraction of meaningful features from the flow table entries, the use of unconstrained statistics makes the system more susceptible to evasion attacks. Furthermore, the reliance on easily manipulated features further exacerbates this vulnerability.

\textbf{TAKAWAY.} Ensembles of DL models exhibit excellent classification performance across a wide range of settings. The ability of DNNs to learn powerful, descriptive features --- even from unlabeled data --- and their capacity to generalize to unseen data make DL-NIDS highly versatile. Furthermore, utilizing ensembles of heterogeneous DL architectures allows for the effective exploitation of both spatial and temporal features, leveraging the specialized strengths of each architecture. However, the opaque nature of the feature learning and decision-making processes of DNNs reduces explainability, making it difficult to interpret how classification decisions are made. Moreover, directly applying DNNs to network-level features without incorporating architectural or data-level safeguards leaves the models highly vulnerable to adversarial attacks, as malicious inputs can be generated through a variety of means to exploit DNN's weaknesses.

\subsection{Discussion}

Extensive research has demonstrated the potential of both traditional ML and DL approaches in NIDS. Traditional ML approaches typically rely on expert knowledge for feature selection and engineering, which simplifies the learning tasks and can improve performance in specific settings. Moreover, their simpler nature allows for better explainability and improved confidence in classification results. On the other hand, DL-NIDS can autonomously learn complex, non-linear relations in the input data and typically generalize better to unseen queries compared to ML-NIDS. 
However, regardless of the specific learning algorithm used, robustness and resilience to adversarial attacks remain insufficiently addressed concerns. DNNs are heavily susceptible to gradient-based attacks such as adversarial examples (see Section~\ref{sec:adversarial_examples}). ML-NIDS are typically non-differentiable and, therefore, may appear robust against evasion attacks. However, adversaries can leverage differentiable approximations of the models to craft adversarial examples, and several specific adversarial attacks have been developed specifically for traditional ML techniques~\cite{calzavara2020treant, zhang2020decision, de2022evading}. Although relying on ensembles generally enhances robustness against adversarial attacks, the use of multiple models significantly increases computational costs and classification time, limiting the practicality of these approaches in certain scenarios.

Ultimately, to develop NIDS that are resilient to adversarial attacks, it is crucial to thoroughly explore the trade-offs between explainability and modeling capability, as well as between robustness and computational complexity. Moreover, it is essential to understand how the structured nature of network traffic and interdependencies between features affect adversarial attacks. While combinations of ML and DL techniques combine the modeling capabilities of DL with the expert knowledge and explainability of traditional ML, significant gaps remain in the literature, mainly due to insufficient consideration for explainability analysis, real-world scenarios, and adversarial robustness. This raises open questions that remain unexplored:

\textbf{Question 1.} How effective are adversarial attacks against structured data such as network traffic?
Can adversarial attack algorithms effectively account for interdependencies and constraints among features to generate realistic manipulations applicable in real-world scenarios? \hfill

\textbf{Question 2.} How well do ensemble methods resist adversarial attacks in comparison to single-learning approaches and adversarial training methods? Is the increased computational cost justified by the enhanced robustness provided?\hfill

The following sections critically review the literature on the vulnerabilities of intelligent NIDS to adversarial threats. By analyzing how current research replies to these questions, we aim to identify advancements and gaps for developing robust IDS able to counteract evolving adversarial strategies.

\begin{table}
\renewcommand{\tabcolsep}{2.75pt}
\def\arraystretch{1.5}
\scriptsize
\centering
\caption{Comprehensive summary of recent papers for the enhancement of machine learning and deep learning based NIDS}
\begin{adjustbox}{max width=\textwidth}
\begin{tabular}{p{0.5cm}|c|p{2cm}|p{1.1cm}|p{2.5cm}|p{2.5cm}|p{3.5cm}|p{3.5cm}}
\hline
\rowcolor{lightgray}
\textbf{Ref.} & \textbf{Year} & \makecell[c]{\textbf{Feature Selection}}  & \makecell[c]{\textbf{Learning}\\\textbf{Method}} & \makecell[c]{\textbf{Algorithms}} & \makecell[c]{\textbf{Dataset}}  & \makecell[c]{\textbf{Pros}}  & \makecell[c]{\textbf{Cons}}  \\ 
\hline
\cite{chkirbene2020tidcs}& 2020  &  Model-based importance score  & Single learning  & ML (DT \& RF)  & NSL-KDD \& UNSW & Low false positives & Feature selection bias, lack of robust features   \\ 
\hline
\cite{alahmadi2020botection}& 2020 &  Markov Chain  & Single learning & ML (RF)  & ISCX, Stratosphere IPS, CTU-13 & Privacy-preserving, unseen detection & Botnet-only, weak to obfuscation/reordering evasion\\ 
\hline
\cite{mehr2019svm}& 2019 & Manual (entropy) & Single learning  & ML (SVM) & Mininet simulation & Transparent, integration with SDN controller  & DoS-only, easily manipulated features \\ 
\hline
\cite{zhou2018deep}& 2018 & DFEL & Single learning  & DL (DFNN) & NSL-KDD \& UNSW-NB15  & Time-saving applications & Inadequate capacity to reduce data dimensionality\\ 
\hline
\cite{manimurugan2020effective} & 2020 & DBN & Single learning & DL (DBN) & CICIDS 2017 & Effective detection rates, potential for scalability & Evasion through feature deletion \\ 
\hline
\cite{alom2017network} & 2017 & AE & Single learning & DL (DNN) & NSL-KDD, IBM NSCS \& IBM TrueNorth & Potential for scalability & Overfitting\\ 
\hline
\cite{laghrissi2021intrusion} & 2021 & PCA+Mutual Information (MI) & Single learning & DL (LSTM) & KDD’99 & Great capability of learning & Overfitting, easily manipulated features\\ 
\hline
\cite{piskozub2021malphase}& 2021 & AE & Single learning & DL (MLP) & ClamAV, Ember, GT Malware, MalRec, MalShare, VirusShare, VX Underground \& Miscellaneous & High detection rate, robustness to noise & Sensitivity to advanced adversarial attacks \\ 
\hline
\cite{kan2021novel}& 2021 & Manual & Single learning & DL (CNN) & Public IoT data set created by \cite{meidan2018n} & Multi-type attacks detection & Biased feature selection\\ 
\hline
\cite{thakkar2023fusion}& 2023  & Sequential Forward Selection (SFS) with statistical ranking & Single learning  & DL (DNN) & NSL-KDD, UNSW-NB-15,  \&  CIC-IDS-2017 &  Informative feature selection & Sensitivity to outliers \\ 
\hline
\cite{marir2018distributed}& 2018  & DBN & Multimodel learning & DL \& ML & KDD’99 CUP, NSL-KDD, UNSW-NB15, \& CICIDS2017 & Robust to imbalanced data & Sensitivity to outliers \\ 
\hline
\cite{imran2022intelligent}& 2022  & Non-symmetric deep AE & Multimodel learning & DL \& ML (stacked (NDAE)\& SVM) & KDD CUP’99  & Less susceptible to overfitting  & Missing subtle features \\ 
\hline
\cite{ravi2022recurrent}& 2022  & Implicit feature learning & Multimodel learning  & DL (RNN, LSTM, \& GRU) \& ML (RF, SVM, \& LR) & KDD-Cup’99, SDN-IoT, UNSW-NB15,  WSN-DS, \& CICIDS-2017 & Automated feature learning, capturing temporal patterns & Overfitting, difficult interpretation \\ 
\hline
\cite{tama2019tse} & 2019 & Hybrid (Evolutionary algorithms+REP Tree) & Multimodel learning & ML (Rotation Forest \& bagging) & NSL-KDD \& UNSW-NB15 & Robust feature selection & Sensitivity to outliers \\ 
\hline
\cite{zhou2020building} & 2020 & Hybrid (CFS+Evolutionary approach) & Multimodel learning  & ML (C4.5, RF, \& Forest by PA) & NSL-KDD, AWID, \& CIC-IDS2017 & Robust feature selection & Sensitivity to feature correlations \\ 
\hline
\cite{gu2019novel} & 2019 & Feature augmentation & Multimodel learning  & ML (DT- EnSMV) & NSL-KDD & Handle the complex feature interdependencies  & Limited feature control \\ 
\hline
\cite{arreche2024two}& 2024 & XAI \& Information gain & Multimodel learning  & DL \& ML (DNN, DT, KNN, MLP, AdaBoost, Logistic Regression, SVM, \& Extreme Gradient Boosting) & RoEduNet-SIMARGL2021, NSL-KDD, \& CICIDS-2017 & Improved individual and ensemble performance & Limited outlier detection \\ 
\hline
\cite{nandanwar2024deep}& 2024 & Implicit feature learning & Multimodel learning  & DL (CNN \& GRU) & N\_BaIoT & Less susceptible to overfitting, simple architecture  & Sensitivity to outliers, missing important features \\ 
\hline
\cite{hasan2022securing}& 2022 & Implicit feature learning & Multimodel learning  & DL (LSTM \& DNN) & N\_BaIoT  &  Scalability and generalizability for handling large datsets & Limited feature control, sensitivity to noisy data  \\ 
\hline
\cite{nazir2024deep}& 2024 & Implicit feature learning & Multimodel learning  & DL (CNN \& LSTM) & N\_BaIoT \& CICIDS2017 & Scalability and generalizability  & Overfitting, sensitivity to outliers \\ 
\hline
\cite{yaser2022improved}& 2022 & AE & Multimodel learning  & DL (CNN \& Bidirectional Long Short-Term Memory) &  ISCX-IDS-2012 \& UNSW2018 &  Improved DDoS attack detection & Overfitting, misinterpretation by CNNs and BDLSTMs, sensitivity to outliers \\ 
\hline
\end{tabular}
\end{adjustbox}
\label{tab:lit_review}
\end{table}

\section{Analysis of Adversarial Threat to Network Intrusion Detection Systems}\label{sec:nids_attacks}
Evasion techniques, such as adversarial examples, present a fundamental threat to intelligent NIDS. However, unlike domains using unstructured data (e.g., image/audio classification, NLP), generating adversarial examples against NIDS is more challenging due to the interdependency and constraints between features in structured data. Network traffic follows specific protocols with predefined features, data types, and relations between different packet fields, making it difficult to manipulate the traffic coherently while also generating effective attacks. Adversarial manipulations must maintain the semantic validity of the data (e.g., changing the protocol field to a non-existent value) and its underlying structure (e.g., changing a packet header format). 
This section presents an analytical study of existing research on methods for generating adversarial examples that evade ML-NIDS, categorized into white-box, black-box, and gray-box attacks (summarized in Table \ref{tab:gen_adv}). By examining the strengths and weaknesses of different approaches, we highlight the threat of adversarial attacks against NIDS and identify significant gaps in the current literature.

\subsection{Generation of White-Box Attacks} 
In white-box attack settings, adversaries have full knowledge of the target NIDS, including architecture and model parameters. While such a level of access might be unlikely in real-world scenarios~\cite{apruzzese2022modeling}, white-box attacks provide valuable insights into potential attack vectors and vulnerabilities.

\subsubsection{Vulnerability of DL-NIDS:} The application of traditional, gradient-based attacks against DL-NIDS has been analyzed by several studies in the literature. In~\cite{mohammadian2023gradient}, the authors analyze the vulnerability of a DNN-based NIDS against JSMA \cite{papernot2016limitations}. The authors explore all possible combinations of network traffic features and employ a modified saliency map to prioritize the features that minimize the true class probability. Effectively, through the modified saliency map, the attack determines the most sensitive feature combination to craft adversarial perturbations. Similarly, in \cite{ayub2020model} Ayub et al. study the vulnerability of MLP NIDS to JSMA under different training dataset. While they highlight the effectiveness of the attack, these studies do not consider the interdependencies and constraints between traffic features. This limits the applicability of the findings, as the authors do not verify whether the adversarial perturbations respect the semantic validity of the data, or whether they affect the expected behavior of the network traffic in a realistic scenario.
 
Pawlicki et al.~\cite{pawlicki2019artificial} explored several adversarial attacks such as CW~\cite{carlini2017towards}, FGSM~\cite{goodfellow2015explaining}, BIM~\cite{kurakin2018adversarial}, and PGD~\cite{madry2017towards} against DNN-NIDS. All the adversarial attacks were grouped into a single attack class for the IDS training and testing. The results of the study highlight the effectiveness of different gradient-based adversarial attacks in causing misclassification. However, since the adversarial attacks were applied without modification, the generated adversarial perturbations did not respect the semantic validity of the network traffic (e.g., negative packet size). The authors work around this issue by zeroing out invalid fields. However, such an approach still does not guarantee the semantic validity of network packets, potentially resulting in perturbed features that cannot be achieved in a real network setting. Furthermore, the authors do not discuss feature interdependencies and do not consider feature constraints in their attacks. In~\cite{hashemi2019towards}, Hashemi et al. study how correlations between network traffic features impact the effectiveness of adversarial attacks. The authors evaluate adversarial attacks against two types of DL-NIDS: 
\begin{itemize} [topsep=0pt]
    \item \textit{Packet-based NIDS:} The proposed attack relies on a local copy of the target NIDS to identify packet-level perturbations that lead to misclassification. The authors define a set of legitimate packet transformations that maintain the semantic validity of the data. Differently from previous works, the authors do not utilize gradient-based techniques but rely on splitting, delaying, and injecting packets in the network.
    \item \textit{Flow-based NIDS:} To account for feature interdependencies and constraints at the flow level, the authors categorize the features into four groups based on how freely each feature can be perturbed while preserving consistency. During the attack, three distinct masks are employed to determine which group of features to alter. The attack proceeds through multiple iterations of perturbations, progressively expanding the permissible range of alterations to generate the adversarial examples.  
\end{itemize}
Searching for an adversarial perturbation with a lower score can create a buffer zone between perturbations and real-world traffic, especially for dependent features outside the attacker's control. However, variations in real-world attack traffic might cause the NIDS score to fall outside this buffer zone, leading to detection. This highlights the challenge of accurately simulating the dynamic properties of real-world network traffic.

\subsubsection{Vulnerability of Traditional ML-NIDS:} Several attacks have been proposed against Traditional ML-NIDS. Anthi et al. \cite{anthi2021adversarial} studied the effectiveness of JSMA against ML-NIDS based on RF and J48 classifiers in ICS environments. Since RF and J48 are non-differentiable algorithms and JSMA requires gradient computation, the authors rely on a pre-trained DNN that is used as a proxy to generate the adversarial perturbations instead of the target the ML-NIDS. While utilizing a differentiable proxy model to generate the adversarial attacks shows promising results, the authors do not address concerns related to the transferability of the attacks to the original models, nor address the problem of feature consistency and semantic validity of the perturbed samples.

In another study, Alhajjar et al. \cite{alhajjar2021adversarial} investigated the sensitivity of different ML-NIDS (e.g., bagging, gradient boosting, linear discriminant analysis, and quadratic discriminant analysis) to adversarial manipulations. The authors used Genetic Algorithm (GA) and Particle Swarm Optimization (PSO) to iteratively modify malicious traffic data and employed a Generative Adversarial Network (GAN) to generate adversarial samples. The GAN's generator generates adversarial samples, while the discriminator tries to distinguish them from real traffic. To respect feature constraints, the authors relied on expert knowledge to identify features that may be altered without disrupting network functionality. While the proposed approach shows promising results and respects feature constraints, feature interdependencies and semantic validity are not taken into consideration. Moreover, the proposed approach is computationally expensive, limiting its applicability to real-time attacks.

\textbf{TAKEAWAY.} While several studies confirm the vulnerability of intelligent NIDS to white-box adversarial attacks, these works frequently ignore crucial real-world complexities such as feature interdependency, feature constraints, and semantic validity of the perturbed samples, highlighting inconsistencies with real-world threat model settings. Moreover, in the case of ML-NIDS, the transferability of adversarial attacks crafted against proxy models to the original ML-NIDS remains uncertain. Further research is required to fully assess the extent of this threat.

\subsection{Generation of Black-Box Attacks}
\begin{figure*}[t]
  \centering
  \includegraphics[width=.8\textwidth]{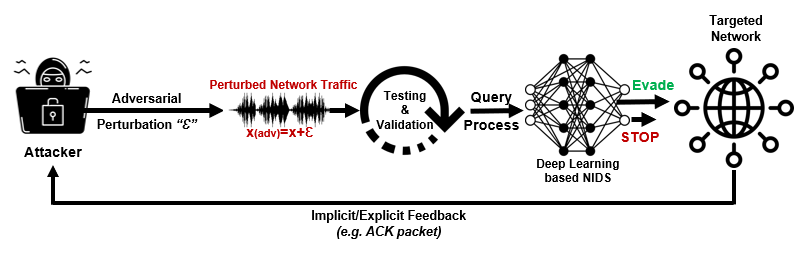}
  \caption{Adversarial perturbation against Neural Network Based-NIDS \textit{(Black-Box)}}
  \label{fig:agent_arch}
\end{figure*}

Black-box attack settings assume that the attacker has no knowledge of the details of the target NIDS, such as architecture or parameters. Compared to white-box attacks, black-box settings provide a more realistic threat model and are more representative of real-world scenarios. Typically, black-box attacks rely on an ``oracle'' model that adversaries can query to validate whether the NIDS detects the perturbed traffic. This feedback from the oracle model is utilized to further refine the perturbation until it achieves evasion. Figure~\ref{fig:agent_arch} illustrates a general framework for these attacks.

\subsubsection{Vulnerability of DL-NIDS:} In \cite{zhang2022adversarial}, the authors evaluate C-LSTM, an ensemble model combining CNN and LSTM, and individual DNN and CNN models against a variety of black-box attacks: Natural Evolution Strategies (NES) \cite{ilyas2018black}, Boundary Attack \cite{brendel2017decision}, HopSkipJumpAttack \cite{chen2020hopskipjumpattack}, Pointwise Attack \cite{schott2018towards}, and OPT-Attack \cite{liu2020query}. The authors assessed various aspects of adversarial sample generation, including attack success rate, query efficiency, and transferability. The results of the study highlight the effectiveness of these attacks against both single-model and ensemble NIDS. However, the study does not properly address concerns related to feature interdependencies and constraints. Moreover, crucial details of the attacks related to feature selection and the query process were not covered, making it difficult to evaluate the real-world applicability of the findings.
In~\cite{peng2019adversarial}, the authors explore the creation of adversarial DoS attacks aimed at evading DNN-NIDS. Their analysis reveals that traditional constraints for generating adversarial examples, which rely on distance metrics such as Lp norms, are ineffective when used for DoS traffic. These metrics treat all features equally, despite DoS traffic containing interconnected features with highly variable value ranges. To address this limitation, the authors utilize Mahalanobis distance, which accounts for feature scale differences, enabling more accurate similarity measures between normal and adversarial DoS traffic. 
While employing different constraints on perturbations helps to better capture individual feature constraints, the approach still overlooks feature interdependencies and semantic validity of the perturbed traffic. Wu et al. \cite{wu2019evading} propose a deep reinforcement learning approach consisting of a black-box botnet NIDS and a flow manipulation module interacting with a Deep Q-Learning agent. 
For each network flow, the agent selects a modification action from a predetermined set that changes communication patterns without affecting functionality. The modified flow is evaluated by the detection model, and the agent is rewarded based on whether or not it is detected as malicious. This feedback loop enables the agent to iteratively refine the learned perturbation strategy and maximize evasion. While relying on a predefined set of modifications ensures that feature constraints and interdependencies are satisfied, it significantly restricts the perturbation space that the agent can explore, thereby limiting the overall effectiveness of the attack. 

In \cite{sharon2022tantra}, the authors propose TANTRA, an evasion attack aimed at models relying on temporal correlations between packets. An LSTM model is used to learn inter-packet arrival times for benign traffic. Then, malicious traffic is generated following the learned pattern, simulating the temporal characteristics of benign traffic without altering the packets' content. While TANTRA can effectively manipulate temporal features to achieve evasion, most NIDS combine timing statistics with other packet or flow information to detect malicious traffic. Consequently, the applicability of this evasion attack is generally limited, as it only targets a subset of the features that are typically used for network intrusion detection.
Debicha et al. \cite{debicha2023adv} focus on evading flow-based NIDS targeted at botnets.
Unlike previous works, the authors train a surrogate model on benign traffic as a proxy for the real NIDS and generate adversarial perturbations relying on the output of the surrogate model. The authors ensure that feature constraints and semantic validity are satisfied through feature masking and projection operations. Feature masking allows perturbation only of features that can be freely modified (e.g., total number of packets), while the projection ensures that the domain of each feature is respected. The attack relies on the transferability of the adversarial attack between the surrogate model and the target NIDS. Similarly, Zolbayar et al. \cite{zolbayar2022generating}, developed NIDSGAN, a GAN-based attack that relies on a surrogate model to generate the adversarial perturbation. The GAN is trained using the surrogate model as a discriminator, and limitations on the samples generated by the GAN are imposed to maintain the semantic validity of the traffic. While using a surrogate model simplifies the generation of the adversarial perturbation, the effectiveness of the attack largely depends on how closely the decision boundary of the surrogate model resembles the original one. Moreover, constraining the features that can be modified reduces the overall perturbation space, potentially limiting the effectiveness of the attack by restricting its ability to find optimal adversarial examples.

\subsubsection{Vulnerability of Traditional ML-NIDS:} Similar to~\cite{zolbayar2022generating}, IDSGAN \cite{lin2022idsgan} utilizes a GAN architecture and a surrogate model to generate adversarial traffic. The surrogate model is employed as a distinguisher for the GAN training, providing a differentiable approximation of the target ML-NIDS. The proposed attack relies on a restricted modification mechanism to ensure that only non-functional features are altered for each type of attack, guaranteeing feature constraints and semantic validity. While the evaluation of the approach shows promising results, it is based on the simple and outdated NSL-KDD dataset, which does not accurately reflect modern real-world network traffic. Furthermore, limiting the allowed perturbations to non-functional features further restricts the attack's effectiveness.

\textbf{TAKEAWAY.} Existing studies suggest that Black-box attacks pose a significant threat to NIDS, regardless of their underlying architecture. However, the full extent of this threat remains uncertain, as many studies fail to account for real-world limitations in their threat models. Furthermore, most studies only partially address feature constraints and semantic validity of the perturbed traffic. These limitations underscore the need for further research to enhance NIDS defenses against evolving black-box attack strategies.

\subsection{Generation of Gray-Box Attacks}
In gray-box threat models, the attacker has partial knowledge of the NIDS features and architecture, allowing adversarial manipulations that are better targeted against the specific NIDS compared to the black-box setting.

Assuming the attacker knows the feature set used by the target NIDS, Vitorino et al. \cite{vitorino2022adaptative} propose the Adaptive Perturbation Pattern Method (A2PM). A2PM generates adversarial examples by analyzing features to identify valid ranges and correlations. It applies interval patterns to perturb numerical features within limits and combination patterns to preserve coherence in categorical features. These patterns are sequenced to introduce perturbations that comply with domain rules and class-specific requirements. 
Sheastsley et al. \cite{sheatsley2022adversarial} propose two approaches to generate adversarial attacks that can adapt to specific network constraints: 
\begin{itemize} [topsep=0pt]
    \item \textit{Adaptive JSMA (ASJMA):} AJSMA incorporates network traffic constraints into JSMA. The attack relies on the concept of primary features, which are features that, when set, limit the permissible range of other features. When generating perturbations, AJSMA verifies whether a given feature is affected by another primary feature and limits the allowed perturbation according to their specific interdependencies.
    \item \textit{Histogram Sketch Generation (HSG):} HSG learns meta-perturbations, called \textit{sketches}, from a set of previously generated perturbations. HSG utilizes the results of previous attacks to detect frequently perturbed features and their modifications to learn a generalized attack pattern. The learned sketch can then be applied to new network data to evade detection.
\end{itemize}
While these techniques result in perturbations that satisfy feature constraints and interdependencies, they require extensive knowledge of features and target network traffic distribution, which may be limiting in real-life settings.
In \cite{jmila2022adversarial}, the authors focus only on evaluating the robustness of ML-based ensembles against various well-known adversarial attacks. To enable the computation of gradient-based attacks on non-differentiable ML models, the authors employ a surrogate DNN model. The adversarial perturbation is computed using the surrogate model, with the success of the attack depending on the transferability of the generated perturbations to the target NIDS. While this paper approved the vulnerability of shallow models to adversarial samples, the original accuracy of the latter has not significantly dropped compared to other existing attacks (e.g., \cite{alhajjar2021adversarial}). Additionally, further information about how the methodology considers the attacker's knowledge of the feature constraints or the data collected during the attack is not provided. Hence, this study does not reflect real-world conditions. 

\textbf{TAKEAWAY.} Gray-box attacks strike a balance between the high complexity of black-box attacks and the limited real-world applicability of white-box attacks, posing a significant threat to intelligent NIDS. However, questions remain about their effectiveness in practical settings and the challenge of balancing attacker knowledge with realistic adversarial perturbations. Further research is required to accurately assess the risk posed by grey-box attacks and develop robust countermeasures for NIDS.

\subsection{Discussion}
This analytical review highlights significant gaps in the literature concerning the practical applicability of proposed adversarial approaches in network traffic. Several studies fail to account for feature constraints and interdependencies in structured data like network traffic. Furthermore, methodological details on how the attacker can gather relevant information for attacks, such as how to obtain a suitable surrogate model in black-box settings, are often lacking. Finally, several black-box proposals heavily rely on oracle access to the target NIDS, which is an unlikely proposition in real-world settings. These limitations underscore a significant gap between theoretical analysis and recommendations, and their practical feasibility.

\begin{figure}[t]
  \centering
  \begin{adjustbox}{max width=\textwidth}
    \includegraphics{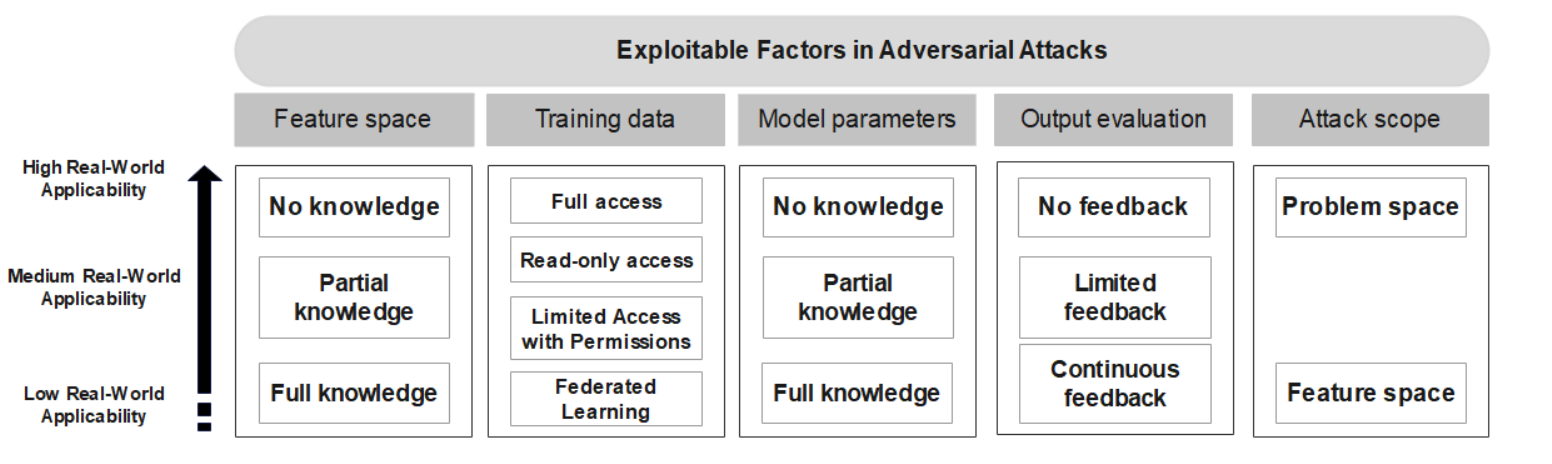}
  \end{adjustbox}
  \caption{Real-world applicability of adversarial factors when generating adversarial attacks.}
  \label{fig:rw}
\end{figure} 

The practical effectiveness of adversarial attacks depends on the attacker's access and knowledge of several key factors~\cite{apruzzese2022modeling}, as highlighted in Figure~\ref{fig:rw}:
\begin{itemize} [topsep=0pt]
    \item \textbf{Feature space:} What level of knowledge does the attacker possess regarding the features utilized by the NIDS for classification?
    \item \textbf{Training data Access:} Is the training data for the ML-NIDS readable, modifiable, or completely inaccessible by the attacker?
    \item \textbf{Model parameters:} What level of knowledge does the attacker possess regarding the model architecture and parameters used by the NIDS?
    \item \textbf{Output evaluation:} Does the attacker have oracle access to the target NIDS, and what restrictions exist on the number of queries they can make?
    \item \textbf{Attack scope:} In what space are the adversarial perturbations generated? In the problem space (network traffic) or in the features space (learned model features)?
\end{itemize}

Further research in the area of robust NIDS should carefully account for the specific type of network being targeted, as well as the expected level of access and knowledge available to the adversary. The threat model considered should be designed to reflect these considerations, and the adversarial attacks used in the analysis must align with the level of access outlined in the threat model.

\begin{table}
\renewcommand{\tabcolsep}{2.55pt}
\def\arraystretch{1.2}
\scriptsize
\centering
\caption{Summary of adversarial attacks on NIDS}
\begin{adjustbox}{max width=\textwidth}
\begin{tabular}{c|c|c|c|c|c|c|c}
\hline

\rowcolor{lightgray}
\multirow{-1}{*}{\textbf{References}} & \multirow{-1}{*}{\textbf{Year}} & \multicolumn{3}{c|}{\textbf{Type of Attack}} & \multicolumn{2}{c|}{\textbf{Applicability}} & \multicolumn{1}{c}{\multirow{-1}{*}{\textbf{Limitation}}} \\ 
\cline{3-7}

\rowcolor{lightgray}
 & & \textbf{White-Box} & \multicolumn{1}{c|}{\textbf{Black-box}} & \multicolumn{1}{c|}{\textbf{Gray-Box}}  & \multicolumn{1}{c|}{\textbf{Target Model}} & \multicolumn{1}{c|}{\textbf{Dataset}}  &  \\ \hline
 
\cite{mohammadian2023gradient} & 2023 & \cmark & \xmark & \xmark & \multicolumn{1}{c|}{\multirow{2}{*}{MLP}} & 
\begin{tabular}[c]{@{}c@{}}CIC-IDS2017, \\ CIC-IDS2018, \\ CIC-DDoS2019
\end{tabular} 
& \begin{tabular}[c]{@{}c@{}}No consideration of feature interdependency, lack of realism
\end{tabular} \\ \hline

\cite{ayub2020model} & 2020 & \cmark & \xmark & \textbf{\xmark}       & \multicolumn{1}{c|}{MLP} & \begin{tabular}[c]{@{}l@{}}CICIDS 2017,\\ TRAbID 2017\end{tabular} & \begin{tabular}[c]{@{}l@{}}No consideration of feature interdependency, lack of realism 
\end{tabular} \\ \hline

\cite{pawlicki2020defending} & 2020  & \cmark & \multicolumn{1}{c|}{\textbf{\xmark}} & \textbf{\xmark} & \multicolumn{1}{c|}{DNN} & CICIDS 2017 & \begin{tabular}[c] {@{}c@{}} No consideration of feature interdependency \\ Introduction of negative values into certain network traffic features 
\end{tabular} \\ \hline

\cite{hashemi2019towards} & 2019 & \cmark  & \multicolumn{1}{c|}{\textbf{\xmark}} & \textbf{\xmark} & \multicolumn{1}{c|}{DNN} & CICIDS 2017 & Generalizability and lack of realism
\\ \hline

\cite{anthi2021adversarial} & 2021  & \cmark  & \multicolumn{1}{c|}{\textbf{\xmark}}  & \textbf{\xmark}  & \multicolumn{1}{c|}{RF, J48}  & Power System & \begin{tabular}[c] {@{}c@{}} Lack of details on handling feature correlations \\ Limited generalizability due to pre-trained model \\ Difficulty in balancing feature modification
\end{tabular} \\ \hline

\cite{alhajjar2021adversarial} & 2021  & \centering \cmark & \multicolumn{1}{c|}{\textbf{\xmark}} & \textbf{\xmark} & \makecell[c]{SVM, RF, DT, \\ NB, MLP, KNN,\\ LR, GB, Bagging, \\ LDA, QDA} & \begin{tabular}[c]{@{}c@{}}NSL-KDD, \\ UNSW-NB15\end{tabular}  & Impact on real-time attacks \\ \hline

\cite{zhang2022adversarial} & 2022  & \multicolumn{1}{c|}{\textbf{\xmark}}  & \cmark & \textbf{\xmark} & \multicolumn{1}{c|}{C-LSTM,MLP, CNN}  & CSE-CICIDS 2018  &   \begin{tabular}[c] {@{}c@{}} No explanation on feature selection and correlations \\ Limited information on the attacker's query process \end{tabular} \\ \hline

\cite{peng2019adversarial} & 2019 & \multicolumn{1}{c|}{\textbf{\xmark}} & \cmark & \textbf{\xmark}  & \multicolumn{1}{c|}{DNN}  & \begin{tabular}[c]{@{}c@{}}KDDcup99,\\ CICIDS2017\end{tabular}  &  \begin{tabular}[c] {@{}l@{}} Limited effectiveness of traditional distance metrics \\ Lack of details on the optimal settings for Mahalanobis distance parameters \\ Lacks of details on how the attacker gathers information about the target model
\end{tabular} \\ \hline 

\cite{wu2019evading}  & 2019 & \multicolumn{1}{c|}{\textbf{\xmark}} & \cmark & \textbf{\xmark}  & \multicolumn{1}{c|}{CNN \& DT}  & \begin{tabular}[c]{@{}c@{}}CTU13\\ \end{tabular}  & \begin{tabular}[c] {@{}l@{}} Limited action space exploration \\ Inability to directly exploit feature correlations \end{tabular}  \\ \hline

\cite{debicha2023adv} & 2023  & \multicolumn{1}{c|}{\textbf{\xmark}} & \cmark  & \textbf{\xmark} & \multicolumn{1}{c|}{DNN}  & Ad-Hoc & \begin{tabular}[c] {@{}c@{}} Subtle, detectable perturbations \\ Lack of explanation on how the attacker gathers information  \end{tabular} \\ \hline

\cite{zolbayar2022generating} & 2022  & \cmark  & \cmark & \textbf{\xmark} & \multicolumn{1}{c|}{DNN} & \begin{tabular}[c]{@{}l@{}}NSL-KDD,\\ CICIDS 2017\end{tabular}  & Using limited data for the surrogate model, reduce the attack success rates  \\ \hline

\cite{lin2022idsgan} & 2022  & \multicolumn{1}{c|}{\textbf{\xmark}} & \cmark  & \textbf{\xmark}  & \multicolumn{1}{c|}{SVM, MLP, KNN, RF, LR} & NSL-KDD & Outdated data, lack of realism\\ \hline

\cite{sharon2022tantra} & 2022  & \multicolumn{1}{c|}{\textbf{\xmark}} & \cmark  & \textbf{\xmark}      & \multicolumn{1}{c|}{LSTM} & \begin{tabular}[c]{@{}c@{}}CIC-IDS2017,\\ Kitsune\end{tabular} & Complex real-time setup (less flexible) \\ \hline

\cite{vitorino2022adaptative} & 2022  & \multicolumn{1}{c|}{\textbf{\xmark}} & \multicolumn{1}{c|}{\textbf{\xmark}}  & \cmark & \multicolumn{1}{c|}{MLP, RF}  & \begin{tabular}[c]{@{}c@{}}CIC-IDS2017,\\ IoT-23\end{tabular}  &  \begin{tabular}[c] {@{}c@{}} If the training data used by A2PM is inadequate, the generated attacks will be \\ unrealistic and detected\end{tabular} \\ \hline

\cite{jmila2022adversarial} & 2022 & \cmark & \cmark  & \cmark & \makecell[c]{AdaBoost, Bagging, \\Gradient Boosting, LR, \\ DT, RF, SVC} & \makecell[c]{NSL-KDD,\\ UNSW-NB 15} & No consideration of feature interdependency, lack of realism\\ \hline

\cite{sheatsley2022adversarial} & 2022   & \cmark & \multicolumn{1}{c|}{\textbf{\xmark}} & \cmark & \multicolumn{1}{c|}{ML \textit{(Not Specified)}}  & \makecell[c]{NSL-KDD,\\ UNSW-NB 15} & Outdated data, lack of realism\\ \hline
\end{tabular}
\end{adjustbox}
\label{tab:gen_adv}
\end{table}

\section{Analysis of Proposed Adversarial Defenses} \label{sec:defenses}

Adversarial attacks pose significant threats to intelligent NIDS, and, to date, defenses are still limited. This section reviews existing countermeasures, which can be mainly categorized into Adversarial Training and Architecture Reinforcing. Table~\ref{tab:def_adv} summarizes the strengths and limitations of the reviewed defenses from the literature.

\subsection{Adversarial Training} 
Adversarial training~\cite{carlini2019evaluating} is a method used to enhance the robustness of machine learning models by including adversarial examples, inputs deliberately altered with small, often imperceptible changes aimed at misleading the model—during the training process. These examples challenge the model by exposing its vulnerabilities, forcing it to learn from both standard and adversarial inputs. As a result, the model becomes more resilient to adversarial attacks, reducing the likelihood of being fooled by manipulated data and improving its overall generalization and performance in more complex, real-world scenarios.

Alahmed et al.~\cite{alahmed2022mitigation} propose an adversarial defense method that utilizes Random Forest classifiers for NIDS. Their approach includes the training of a Generative Adversarial Network (GAN) to generate realistic adversarial network traffic examples. The NIDS model is then trained in two phases: first on real traffic data, and subsequently on a combined dataset that includes both the adversarial examples produced by the GAN and the original data. This two-step training process bolsters the model's defenses against adversarial perturbations. The approach is further enhanced through Principal Component Analysis (PCA) for feature selection, and the model’s resistance is evaluated using a black-box attack technique.  Despite its strengths, this approach suffers from various drawbacks and limitations. Firstly, the reliance on GANs to produce synthetic network data is costly and inefficient. GANs are shown to be tricky to train and stabilize. Secondly, and more crucially, the GAN-generated network data are not validated as real network traffic flow data. The fact that GANs are able to generate data from a given distribution does not mean that the generated data represents real-world traffic flows.
Similarly, Debicha et al~\cite{debicha2021adversarial} aim at strengthening a DL-based NIDS through the use of adversarial training. To generate the adversarial examples, they rely on three off-the-shelf adversarial generation techniques, namely the FGSM, PGD, and BIM. Similarly to the previous approach, the proposed technique does not consider the constraints required for the generated adversarial network flow to be realistic. Additionally, it shares the same limitation of having limited practical application.
Venturi et al. ~\cite{venturi2023arganids} address network intrusion detection by monitoring network traffic through a graph representation, including both topology and individual connections. The approach uses a Graph Neural Network (GNN) within an AE to condense graph complexities into embeddings that capture important components of each flow and their interactions, acting as a feature extractor. To improve resilience against data perturbations, they rely on adversarial training as well, exposing the AE to both legitimate and malicious traffic to learn robust embeddings. These embeddings, containing network architecture and flow details, are then processed by an RF classifier for normal and abnormal traffic classification. A key limitation to this procedure is its high computational cost, especially in resource-constrained environments; therefore, optimizing this process for greater efficiency represents a viable direction for building adversarially robust NIDS.
Another approach, the EIFDAA framework~\cite{li2023eifdaa}, provides active defense through adversarial training based on three steps. First, a simulation of attacks in which adversarial examples are generated using BIM, PGD, FGSM, and DeepFool. These attacks modify legitimate data to create malicious samples that mimic typical traffic patterns and evade IDS detection. The second step is adversarial evaluation, which requires including original attack data that represents real attacks together with simulated attacks into the IDS being evaluated. A significant decrease in attack detection indicates an IDS vulnerability. Finally, adversarial training allows the IDS to learn attacker techniques and fortify its defenses against future attempts to exploit similar vulnerabilities by retraining it using a combination of adversarial examples and the original training data. EIFDAA offers a comprehensive evaluation and defensive method. However, its performance depends on the accuracy of the evaluation data and the quality of attack simulations. 

Recently, Debicha et al.~\cite{debicha2023adv}, aimed at crafting a filter-like system to prevent adversarial examples from reaching and fooling the NIDS. This detector acts as a pre-filter using three techniques: adversarial training, adversarial detection, and ensemble methods. Three MLPs are trained in parallel; each one focuses on a specific set of network traffic features based on the possibility of manipulating them by the attacker (modifiable or unmodifiable). The training data is composed of two classes: adversarial and clean. During detection, each MLP monitors an incoming data point. Each prediction is then weighted based on its training accuracy. The weighted predictions from all three MLPs are integrated using a combination of contextual discounting and Bayesian fusion, leading to a final classification of each data point as either \textit{adversarial} or \textit{clean}. This classification determines whether the data point is permitted to pass through to the NIDS. While effective, running multiple models can be computationally expensive, and accurately classifying features may be difficult due to manipulation by attackers.

\textbf{TAKEAWAY.} 
Adversarial training has become one of the most widely used technique for enhancing the robustness of machine learning models, particularly DNNs. This is due to its ease of use, its flexibility, and its demonstrated effectiveness in improving model resilience. However, several considerations must be taken into account when implementing adversarial training within ML pipelines.
Firstly, there is a risk of overfitting. Adversarial training alters the decision boundary learned by the model, with the goal of making it more robust to malicious outliers (adversarial examples). However, incorporating an excessive number of adversarial examples or allowing overly large perturbations poses the risk of excessively altering the decision boundary, which can lead to misclassifications and diminished generalization ability. Secondly, and most importantly, the adversarial sample generation techniques must be carefully selected and tuned to ensure that the generated samples are meaningful representations of actual network traffic, relevant to the task of NIDS. Particularly, to maximize the robustness of the model, it is crucial to consider feature constraints and interdependencies during adversarial example generation.

\subsection{Architecture Reinforcing} 
A novel area of research in defending against adversarial attacks is architecture reinforcing ~\cite{9156305,hitaj2023minerva}. This field focuses on developing neural network architectures and designing models that are inherently robust to adversarial attacks by design.

Jiang et al.~\cite{jiang2022fgmd} rely on multi-model and task modelling to achieve adversarial robustness. They analyze network data and divide features into three groups: time-related, packet-related, and state-related, which are less likely to be altered by attackers. Each group is processed by a separate DL model, and the predictions are combined to decide if a flow is malicious. This approach shows good adversarial robustness, while maintaining adequate detection rates in the absence of adversarial attacks. However, its effectiveness depends on accurately predicting which features attackers will target, which can reduce its reliability when adversaries alter their attack. 

Jmila et al.~\cite{jmila2022adversarial} assessed the robustness of Gaussian data augmentation against adversarial attacks by introducing random Gaussian noise into the training data. This method aims to expand the dataset and limit the ability of attackers to craft targeted adversarial examples, making the IDS less vulnerable to feature manipulation. The authors compared IDS classifiers trained on original versus augmented data, evaluating them on both normal and adversarial test sets. They measured the differences in accuracy and False Negative Rate to assess the model's performance. While this approach enhances IDS resilience, its effectiveness may be constrained by the inability to fully simulate real-world adversarial manipulations.

Zhang et al.~\cite{zhang2020tiki} propose another method that incorporates three strategies to defend against different types of black-box adversarial attacks, namely ensemble voting, ensemble adversarial training and adversarial sample detection, with each strategy having its limitations. For instance, combining multiple deep learning models can increase complexity and heighten the risk of misclassification. Adversarial training, while useful, relies heavily on the attacker’s knowledge and risks overfitting. Query detection requires careful threshold tuning and remains vulnerable to attacks that can bypass the initial detection phase. Additionally, these methods raise computational costs and may struggle to keep pace with rapidly evolving threats.

In contrast to traditional models that explicitly classify attack types, a two-layer hierarchical defense system was proposed by Mccarthy et al.~\cite{mccarthy2023defending}. It uses a coarse layer to group similar attacks (e.g., various DoS attacks) and a fine layer to classify specific types (e.g., Slowloris). This makes it challenging for attackers to inject perturbations that trick both layers. Despite the high computational cost of this approach and the need to define the attack group hierarchy, it results in a robust model that prevents adversarial examples and maintains high classification accuracy in regular traffic.

Recently, Mutation-based detection has proven effective for adversary defense~\cite{he2023adversarial} as have approaches that rely on observing adversarial manipulations occurring near the decision boundary~\cite{stutz2019disentangling}. Wang et al.~\cite{wang2022manda} identify adversarial perturbations by detecting inconsistencies between the NIDS's classification and natural data patterns (manifolds). These manifolds represent the original data structure and decision boundary, making it hard to detect perturbations as they share characteristics with both malicious and benign traffic. The effectiveness of this approach relies on the accuracy of the IDS model, leaving it vulnerable to exploitation if the model has any weaknesses. Moreover, this dual approach, involving manifold analysis and decision boundary detection, is more computationally expensive compared to previous detection techniques.

\textbf{TAKEAWAY.} Rather than focusing on specific attack patterns, the architecture reinforcing of NIDS aims to develop more generalized and adaptable defenses. This broader approach enhances the system’s resilience to a wide range of emerging threats, reducing the need for frequent updates to address new attack types. Additionally, this strategy simplifies model maintenance and reduces the need for constant retraining. However, it requires careful consideration of trade-offs between security, accuracy, and complexity. Striking the right balance is essential to ensure the system effectively counters evolving threats without becoming excessively resource-intensive or cumbersome. Furthermore, differently from adversarial training, robust architectures need to be designed specifically for a target task and setting, limiting their versatility.

\subsection{Discussion}
Despite the rapid advancements in adversarial machine learning, there remains a substantial gap in the development and implementation of effective defenses for NIDS. 
Current research predominantly emphasizes the analysis of adversarial attacks, with comparatively less focus on devising practical and resilient defense mechanisms. Adversarial training has emerged as a promising approach for enhancing the robustness of NIDS against adversarial examples. However, its effectiveness heavily relies on the quality and diversity of the training data, making the process resource-intensive. Furthermore, excessive use of adversarial training may lead to overfitting and reduced generalization capabilites. Alternatively, approaches leveraging robust architectures such as ensembles methods or robust loss functions have been proposed to improve resistance to adversarial manipulation. Nevertheless, these methods are often inadequate in addressing the continuous evolution of adversarial attacks. Hybrid strategies integrating adversarial training with robust architectures may lead to more robust NIDS solutions. However, the current lack of understanding regarding the root causes of adversarial examples, coupled with the limited research on robustness in structured data domains, complicates the identification of potential solutions. Given the dynamic and evolving nature of adversarial threats and the peculiar characteristics of structured data, further research is required to identify promising defense mechanisms to improve the robustness of NIDS.

\begin{table*}[t]
\renewcommand{\tabcolsep}{2.55pt}
\def\arraystretch{1.2}
\definecolor{lightgray}{gray}{0.82}
\scriptsize
\centering
\caption{Comprehensive summary: Analytical study of adversarial defense strategies to NIDS}
\begin{adjustbox}{max width=\textwidth}
\begin{tabular}{c|c|c|c|c}
\hline
\rowcolor{lightgray}

\textbf{Reference}    & \textbf{Year}             & \textbf{Defense Method}   & \textbf{Strengths}  & \textbf{Limitation}                  \\ \hline

\cite{alahmed2022mitigation}  & 2022 & \multirow{3}{*}{\textbf{Adversarial Training}} & \begin{tabular}[c]{@{}c@{}}High detection rate of FGSM perturbations,\\  while maintining the original detection level\end{tabular}   &  High training cost and limited generalizability    \\ \cline{1-2} \cline{4-5} 

\cite{debicha2021adversarial} & 2021  &     & Detection for some specific adversarial attacks  & Reliance on outdated data and minimal adversarial training                                          \\ \cline{1-2} \cline{4-5} 

\cite{venturi2023arganids} & 2023  &     & Strong against specific adversarial attacks  &  Computational cost, limited generalizability, and limited adversarial training                                    \\ \cline{1-2} \cline{4-5} 

\cite{li2023eifdaa} & 2023  &     &  Strong against some specific adversarial attack  & Limited generalizability and reliance on accurate data  \\ \cline{1-2} \cline{4-5} 

\cite{debicha2023adv}   & 2023   &    & High detection performance & Computationally expensive, vulnerable to evolving attacks
\\ \hline

\cite{jiang2022fgmd}    & 2022  & \multirow{6}{*}{\textbf{Architecture Reinforcing}} & Improved attack detection through multi-modal analysis  & Limited robustness to adaptive and evolving attacks                    \\ \cline{1-2} \cline{4-5} 

\cite{jmila2022adversarial}   & 2022  &     &  Simplicity and improved generalizability & Loss of information, restricted to specific attacks 
\\ \cline{1-2} \cline{4-5} 

\cite{zhang2020tiki}   & 2020   &  & High detection performance & Computationally expensive and limited adaptability                     \\ \cline{1-2} \cline{4-5} 

\cite{mccarthy2023defending}  & 2023   &  & Strong robustness to evasion  & Computationally expensive \\ \cline{1-2} \cline{4-5} 

\cite{wang2022manda}   & 2022  &   & High true positive rate with low false positives & \begin{tabular}[c]{@{}c@{}}Computationally expensive\\ Its effectiveness depends on the accuracy of the target model \\ Not adaptive to more sophisticated attacks \end{tabular}                                                  \\ \hline
\end{tabular}
\label{tab:def_adv}
\end{adjustbox}
\end{table*}

\section{Lessons Learned, New Insights and Guidelines for Future Research}
\label{sec:lessons_learned}
This section explores lessons learned and potential future research directions, providing insights and guidelines to improve future research in the area.

\subsection{Inter-dependency of Structured Features}
As discussed in Section~\ref{sec:nids_attacks}, existing studies often fail to consider the potential impact of feature constraints and correlations when designing adversarial attacks against NIDS. This leads to a gap between theoretical analysis and recommendations, and their practical feasibility. Key characteristics that require consideration include: 

\begin{itemize}
\item \textbf{Syntactic and Semantic Validity:} The attack might be invalid if packet headers or content are modified without considering the context of the network communication. Network protocols have well-defined syntactic and semantic rules that dictate the structure and behavior of traffic. Arbitrary modifications to packet data may lead to violations of these protocols.
\item \textbf{Time-based dependencies:} Network traffic often presents time-sensitive patterns. If adversarial perturbations incorrectly alter these temporal dependencies (e.g., sending an acknowledgment before receiving the associated data packet), it will lead to protocol violations and lead to failure in real-world settings.
\item \textbf{Dependent features:} Modifications to certain attributes, such as packet size or frequency, lead to changes in other features, such as total number of bytes or communication duration. Adversarial attacks and defenses must consider these dependencies to be effective in real-world settings.
\end{itemize}

Practical adversarial attacks against NIDS must consider these characteristics to ensure realistic analysis and results that can translate into real-world settings. Neglecting these factors may lead to findings that are disconnected from real-world scenarios.

\subsection{Lack of Defense Strategies against Adversarial Attacks}
This survey highlights the lack of robust defenses capable of addressing various types of adversarial attacks. Multiple factors contribute to this shortcoming, posing a significant and persistent challenge for ML models:

\begin{itemize}
\item \textbf{Lack of explainability:} DL models often lack interpretability and transparency due to their complex decision-making processes and architectures, making it challenging to understand their internal functions~\cite{hao2023manipulating}. This hinders the development of effective defenses against adversarial attacks, as the models' weaknesses are not fully understood. 

\item \textbf{Dynamic nature of adversarial examples:} Adversarial attacks are constantly evolving as attackers develop new strategies to exploit models' vulnerabilities. This evolution is driven by the open nature of research, and the availability of large model architectures and datasets, which enable attackers to refine their techniques ~\cite{sun2023differential}. Consequently, static defenses face a significant challenge in keeping up with the rapidly changing landscape of these attacks.

\item \textbf{Generalization:} Existing defenses against adversarial examples are developed with specific attack scenarios and datasets, making them difficult to apply to other contexts and real-world settings. Therefore, these countermeasures may not generalize well and might fail against new attack variations on different datasets.
\item \textbf{Transferability:} Adversarial manipulations created for one model can often be successfully transferred to other models, a challenge known as \textit{transferability} \cite{xie2019improving}. Hence, an effective defense for one model may be ineffective for another with similar vulnerabilities against new attack iterations within different datasets.

\end{itemize}
Intelligent models remain susceptible to adversarial attacks, which prevents their safe deployment in cybersecurity. 
This weakness stems from both the complexity of these attacks and the incomplete understanding of how they exploit model flaws. To ensure the safe use of machine learning, further research is needed on defense strategies and the explainability of adversarial attack behaviors.

\subsection{Lack of Real-World Relevance}
The analytical study conducted in Section \ref{sec:nids_attacks} identifies a critical gap in the literature: the lack of real-world applicability. This gap arises from several factors: 
\begin{itemize}
\item \textbf{Dataset limitations:} Some datasets are outdated and/or fail to capture the complexity and diversity of real-world network traffic, which undermines the credibility of experimental findings.

\item \textbf{Simplified attack techniques:} Most existing studies consider complete control over datasets and full knowledge of the target NIDS, which does not reflect real-world conditions.
\item \textbf{Evolving Attack Strategies:}
Researchers frequently focus on established adversarial attack methods, overlooking the adaptability and creativity of real-world attackers, who might exploit previously undiscovered vulnerabilities, compromising the effectiveness of current defense strategies.
\end{itemize}

By addressing these limitations, future research can provide more practical insights to enhance the development of robust defenses against emerging cyber threats.

\subsection{Placement of Adversarial Examples}

\begin{figure}[t]
  \centering
  \includegraphics[width=0.5\textwidth]{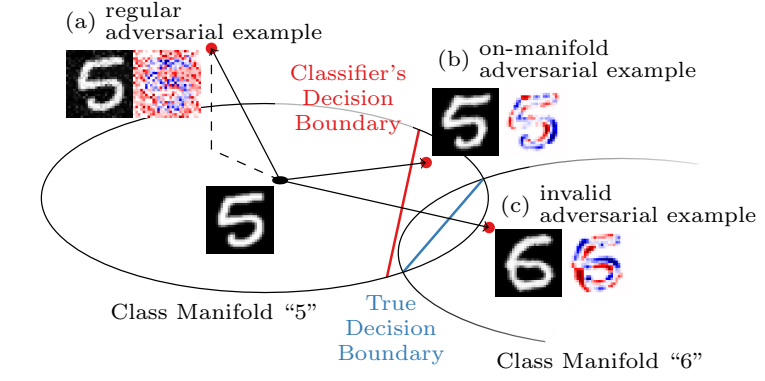}
  \caption{Comparative illustration of standard, on-manifold \textit{(constrained)} adversarial examples vs. unconstrained ones \cite{stutz2019disentangling}.}
  \label{fig:manifold}
\end{figure}

Assuming data can be represented as collections of manifolds~\cite{stutz2019disentangling}, many studies explored the vulnerabilities of adversarial examples. They characterized adversarial subspaces and examined the relationship between adversarial perturbations and manifolds. These investigations can be categorized into the following groups:

\begin{itemize}
    \item \textbf{Leave-the-manifold attacks:} They create perturbations that force input data outside the manifold (or learned data distribution). Researchers (e.g., \cite{tanay2016boundary, jalal2017robust, samangouei2018defense, schott2018towards, gilmer2018relationship}) examined this concept, noting that these modifications are more intense and introduce features not in the training dataset. They are larger and more visible, exploiting the model's weaknesses in handling unfamiliar or out-of-distribution data.
    
    \item \textbf{On-manifold attacks:} Another possible position of adversarial attacks is observing them on-manifold~\cite{stutz2019disentangling}, as it is illustrated in Figure~\ref{fig:manifold}. This concept states that training data and adversarial samples belong to the same distribution or manifold. In other words, these attacks stay within the feature space that the model is familiar with from training. Compared to Leave-the-manifold attacks, these perturbations are usually subtle and do not visibly alter the input data.
    
    \item \textbf{Low-dimensional adversarial subspaces:} This option refers to adversarial examples existing in a lower-dimensional subspace of the feature space \cite{ma2018characterizing}. By exploiting this structure, attackers can find perturbations that significantly affect the model's output with minimal input modifications.
\end{itemize}

\textbf{Insight.} These findings have been extensively explored in the image domain, where the input space is mostly unconstrained and there are soft interdependencies between features. Structured data such as network traffic, however, is typically characterized by hard feature interdependencies and constraints that can significantly affect the space of allowable adversarial examples and their placement. This research gap presents an opportunity to explore a better characterization of adversarial examples in the context of structured data. Current defenses against adversarial attacks used in NIDS are typically based on heuristic approaches that do not necessarily take into consideration the localization of adversarial examples in the domain space. Gaining a deeper understanding of adversarial perturbations and the relationship between feature interdependencies and the positioning of adversarial examples within the network data space can lead to significant improvement in NIDS robustness.

\subsection{Good Practices}
As the volume and complexity of data that must be processed per unit of time continue to increase, systems such as NIDS must utilize techniques like ML for rapid and efficient detection. However, several factors need to be taken into account before implementing ML-based approaches for network intrusion detection:

\begin{itemize}
    \item \textbf{Adversarial training:}  Incorporating adversarial perturbations during the training process of a NIDS inherently increases robustness and resilience against evasion attempts during inference. However the adversarial training should be carefully done by using plausible adversarial data and not just rely on simplistic adversarial generation techniques that would fool the machine learning model but do not represent valid network data. This would allow the model to actually become robust towards attacks that might actually happen, not just overfit on non-plausible adversarial data.
    
    \item \textbf{Ensemble learning and fusion:} In addition to adversarial training, ensemble learning approaches have proven effective against adversarial attacks. Combining multiple diverse architectures, each trained differently, allows them to learn slightly different decision boundaries, improving overall performance and robustness. The independent errors of these models contribute to better performance and increased overall adversarial robustness since not all models are affected similarly by the same adversarial sample.
    Robustness can be further tuned through different model fusion methods (e.g., voting methods, averaging, and meta-learning), leveraging the strengths of individual models. For ensemble-based NIDS, carefully studying the resilience of individual models to several adversarial attacks can help tailor model fusion strategies to enhance adversarial robustness, making NIDS less vulnerable to attacks.

    \item \textbf{Robust feature engineering:} Incorporating high-level semantic features vs. low-level or raw data in NIDS. Incorporating high-level semantic features rather than low-level or raw data in NIDS generally provides a more robust and resilient foundation for machine learning models.
    Focusing on semantic understanding and abstraction improves model stability, generalization, and resistance to adversarial attacks, increasing reliability and effectiveness in real-world applications.

    \item \textbf{Input preprocessing:} 
    Input preprocessing plays a crucial role in enhancing the robustness of machine learning models against adversarial attacks. By applying techniques such as data normalization, feature scaling, noise filtering, and input transformation, preprocessing can reduce the sensitivity of a model to small, malicious perturbations that adversaries introduce to manipulate predictions. These steps help the model focus on the essential patterns within the data rather than being misled by subtle alterations designed to exploit weaknesses. Effective input preprocessing can act as a first line of defense, mitigating the impact of adversarial examples and contributing to a more resilient, reliable system.

    \item \textbf{Anomaly detection:} 
    Anomaly detection can significantly enhance the adversarial robustness of a machine learning model by identifying deviations from expected or normal behavior, thereby flagging potentially malicious inputs. Since adversarial attacks often involve introducing subtle, crafted perturbations designed to mislead the model, anomaly detection techniques can act as a safeguard by detecting these out-of-distribution inputs that do not align with typical data patterns. Regarding ML-based NIDS, envisioning the detection system as a multi-step approach by having an anomaly filter able to detect these suspicious inputs can be a way to augment the overall performance of the system as a whole.

    \item \textbf{Proper (extensive) evaluation:} 
    Proper model evaluation is critical for ensuring the robustness of a machine learning model, particularly in the face of adversarial attacks. When it comes to NIDS, particular care must be given to evaluating and stress testing the model (and the IDS in general) in scenarios pertaining to the threats it is envisioned to withstand. As such the system should be thoroughly evaluated on not just standard network data, which in many cases are outdated, but on timely data that includes also corner cases, out-of-distribution data that fall under the overall threat surface envisioned in the threat model of the system. 

\end{itemize}

\section{Conclusion and Future Directions}
\label{sec:conclusions}
This survey provides a comprehensive and current analysis of adversarial attacks, emphasizing the dynamic landscape of cybersecurity threats, with a specific focus on network intrusion detection systems. It serves as a valuable resource for researchers, organizations, and cybersecurity professionals, enhancing their understanding of this rapidly evolving field. Additionally, the survey details various evasion techniques employed by adversaries to circumvent NIDS, identifies critical gaps in the existing literature, and outlines emerging challenges and insights that can inform future research and development efforts in this field.

\bibliographystyle{ACM-Reference-Format}
\bibliography{main}

\end{document}